\documentclass{article}

\usepackage{arxiv}

\usepackage{amsmath,amssymb,amsfonts}%

\usepackage[utf8]{inputenc} % allow utf-8 input
\usepackage[T1]{fontenc}    % use 8-bit T1 fonts
\usepackage[colorlinks,allcolors=blue]{hyperref}       % hyperlinks
\usepackage{url}            % simple URL typesetting
\usepackage{booktabs}       % professional-quality tables
\usepackage{amsfonts}       % blackboard math symbols
\usepackage{cleveref}       % smart cross-referencing
\usepackage{graphicx}
\usepackage{natbib}
\setcitestyle{numbers,square}
\usepackage{doi}

\usepackage{multirow}%
\usepackage{amsthm}%
\usepackage[title]{appendix}%
\usepackage{xcolor}%
\usepackage{textcomp}%
\usepackage{manyfoot}%
\usepackage{booktabs}%
\usepackage{listings}%
\usepackage{verbatim}
\usepackage{pgfplots}\pgfplotsset{compat=1.18}
\usepackage{subcaption}
\usepackage[version=4]{mhchem}
\usepackage{csquotes}
\usepackage{dcolumn}
\newcolumntype{d}[1]{D{.}{.}{#1}}

\title{Myths of Nuclear Graphite in World War II,\\with Original Translations}

\usepackage{authblk}

\author[1\thanks{Correspondence should be addressed to: \texttt{pjp2136@columbia.edu} or 221 Nassau St., 2nd Flr., Princeton, NJ, 08542, USA.}]{\href{https://orcid.org/0000-0001-7223-1834}{Patrick J. Park}}
\author[2]{Sebastian Herzele}
\author[3]{\href{https://orcid.org/0000-0002-0082-0514}{Timothy W. Koeth}}
\affil[1]{Program on Science and Global Security, Princeton University, USA}
\affil[2]{College of Polymer Engineering, HTL Bregenz, Austria}
\affil[3]{Department of Materials Science and Engineering, University of Maryland, College Park, USA}

\date{}

\renewcommand{\headeright}{P. J. Park et al.}%{Technical Report}
\renewcommand{\undertitle}{}%{Technical Report}
\renewcommand{\shorttitle}{Myths of Nuclear Graphite in WWII}

\hypersetup{
pdftitle={Myths of Nuclear Graphite in WWII},
pdfauthor={Patrick J.~Park, Sebastian Herzele, Timothy W. Koeth},
pdfkeywords={graphite, Uranverein, Heisenberg, nuclear, Bothe},
}

\begin{document}
\maketitle
\vspace{-3em}
\begin{center}
    March 15, 2025
\end{center}
\vspace{2em}
\begin{abstract}
We re-examine a common narrative that experimental errors by Walther Bothe in 1941 led Germany to abandon graphite as a reactor moderator during World War II. Using document-based nuclear archaeology, we first show that both American and German scientists used an incorrect carbon scattering cross section, thereby undermining the accuracy of all wartime data, including their conclusions on carbon's absorption. Moreover, we argue that the availability of exceptionally pure petroleum coke in the United States, rather than any academic breakthrough, decisively enabled their production of nuclear-grade graphite. In contrast, Bothe's Siemens electrographite had more boron contamination than any graphites considered in Fermi's experiments, rendering it genuinely impractical as a moderator. By reframing the decision to eschew graphite as a deliberate decision rather than a mere experimental oversight, we believe the German decision was a rational consequence of a complex interplay between material constraints and wartime priorities. 
\end{abstract}


\section{Introduction}\label{sec1}

The failures of the wartime German nuclear program---dubbed the Uranverein, the ``Uranium Club''---to build a successful nuclear reactor have been well-studied by historians and physicists alike. One particular topic of interest is the Germans' choice to eschew graphite as a possible reactor moderator in favor of heavy water. The oft-repeated narrative is that in January 1941, physicists Walther Bothe and Peter Jensen at the Kaiser Wilhelm Institute for Medical Research (KWImF) tested the absorption of neutrons by graphite. Two mistakes are said to have been made: first, they had air gaps between graphite bricks that caused extra absorption; second, they did not realize boron impurities contributed to absorption within the graphite. These experimental errors ostensibly caused the Uranverein to abandon carbon moderation altogether in February 1942. 

Immediately after the war, Werner Heisenberg published a summary of the Uranverein's progress in \textit{Nature} \cite{heisenberg47}. Their reasons for ruling out graphite were left ambiguous, caused by ``either insufficient consideration of impurities or by deficiencies in the theory.'' As the years progressed, Heisenberg and his immediate collaborators became more blas\'e in shifting their blame to Bothe. When asked in a 1967 interview, ``Why wasn't there more interest in graphite knowing that heavy water was very scarce?,'' Heisenberg responded:

\begin{displayquote}
It was because of the experiment of Bothe's on [neutron absorption in] graphite which was not correct... He had built a pile of graphite pieces but in between the graphite pieces there was always some air and the nitrogen of the air has high neutron absorption. His values were too high but we assumed they were correct and so we did not think carbon could be used.~\cite{ermenc}
\end{displayquote}

Despite work of historian Mark Walker discrediting uncritical acceptances of Heisenberg's claims, the narrative of ``Bothe's mistake'' echoed its way into modern retellings of the nuclear race. A cursory Google search for ``Nazi graphite'' shows a litany of amateurs and professionals, ranging from online forums to the Atomic Heritage Foundation to scholars like P. L. Rose, hyperfocusing on this one experiment as the specious reason for the Germans' failure~\cite{rose,ahf}. Even Richard Rhodes opined, ``...You needed pure graphite. Evidently, the Germans didn’t understand that for reasons I’ve not seen in the record. It seems to have been a mistake.... The graphite measurement may have diverted the German bomb program''~\cite{rhodes16}. % 's Pulitzer Prize-winning account says, ``In January [1941] a misleading measurement... ended German experiments on graphite''~\cite{rhodes12}. Rhodes 
This is perhaps surprising considering how much attention has been given to early nuclear programs. For how central graphite blocks were to Chicago Pile 1's success, very few secondary sources have scrutinized how they were produced. Nor do those accounts seem to fully appreciate just how close the Americans were to ruling out graphite as well. This in turn obfuscates attempts to understand why the Uranverein chose against graphite moderation.

In this paper, we present technical analyses of American and German graphites, as well as a comparative history of their graphite experiments. In doing so, we hope to make three points: First, neither Fermi's nor Bothe's results for neutron absorption in carbon can be taken at face value as both used an incorrect cross section for carbon scattering. Secondly, the availability of very pure petroleum coke with which to make graphite had a profound effect on what reactor designs were pursued. Thirdly and consequently, Bothe correctly concluded the impracticality of German graphites for nuclear purposes. % Ultimately, both programs were largely comparable, as argued by Walker, but only diverged in 1942 with the American discovery of a particularly pure coke in Pennsylvania.

To support our analysis, we append our translations of 13 relevant Uranverein reports, full and excerpted: two from Werner Heisenberg (report codes G-39, G-40), one from Georg Joos (G-46), two from Karl-Heinz H\"ocker (G-41, G-42), three from Walther Bothe (G-12, G-71, and a 1944 journal article), three from Wilhelm Hanle (G-35, G-85, G-153), from the Heereswaffenamt in February 1942, and one from post-war author Wolfgang M\"uller.

\section{Manufacturing Nuclear Graphite}\label{s:2}

We first give an introduction to cross sections and how graphite is produced and purified. This will set the basis for understanding how physicists measured and calculated neutron absorption in graphite.

\subsection{Cross Sections}\label{s:2-xs}
Probabilities of nuclear reactions are represented using cross sections, measured in units of area. Imagine shooting at a barn wall that is 1~m$^2$ in size. While we cannot determine the exact chance of hitting it without more details, if we double the wall's area to 2~m$^2$, the likelihood of a hit also doubles. This illustrates that the relative probability is proportional to the target area. Nuclear physics uses this principle to express reaction probabilities using areas. The symbol for cross sections is \(\sigma\) and its unit is the ``barn,'' where \(1\,\text{b} = 10^{-24}\,\text{cm}^2\).

Neutron energies are grouped into regions of similar behaviors for convenience: thermal (< 1 eV), intermediate (1 eV to 100 keV), and fast (> 100 keV)~\cite{DOE1}. Neutrons interact with nuclei mainly through scattering and absorption, and their cross sections are additive: total cross section = absorption + scattering cross sections. The absorption cross section further separates into fission and capture probabilities: absorption cross section = fission + capture cross sections. Materials with low capture, like heavy water or carbon, must be chosen to sustain the fission reaction. As moderators themselves do not fission, ``capture'' and ``absorption'' are interchangeable terms for them. In this paper, all reported cross sections will be for thermal energies unless explicitly stated otherwise.

\subsection{Graphite Production and Purification} \label{s:2-graphite}

Carbon is found in many forms, but for reactor moderation, the one with highest density is preferred: graphite. Even today, graphite's unique ceramic and metallic properties make it ideal for high-temperature or electrical purposes, of which the purest are used in motor brushes and electrodes, called ``electrode-grade'' or ``electrographite.'' Graphite can technically be made from any organic material, like sugar, but the only economical option in tonnage quantities is by using coke, a byproduct of heating fossil fuels. 
Petroleum coke (petcoke) allows the purest product in sufficient quantities compared to those derived from coal, like metallurgical or tar-pitch coke~\cite{macpherson55}. So, electrographite is often made with petcoke, while coal-derived feeds are used when petroleum is limited or for lower grades of graphite.

To make graphite, coke is compacted and---after other steps---heated to about 2,400$^\circ$C in an airless, electrical furnace. This temperature is where carbon graphitizes, i.e., crystallize into graphene sheets. Resistive heating is used as carbon autoignites in oxygen at 700$^\circ$C~\cite{eatherly62}. Paradoxically, petcoke's purity slows down its heating rate as carbon is a conductor, so metcoke is lined to add resistance. This bleeds back in tiny impurities that affect its nuclear properties~\cite{hilberry65}. 

% When resistively heating petcoke, its purity works against it as carbon is a conductor, so metcoke is added  % When using petcoke, its purity works against it during graphitization, as carbon is a conductor, so it is often lined with metcoke to add electrical resistance. This bleeds back in tiny impurities that affect its nuclear properties~\cite{hilberry65}. 

Manufacturers gauge impurities in graphite by its ``percent ash'': the graphite is purposefully incinerated in air, leaving behind everything not carbon as ``ash,'' which is reported as a fraction of the original mass. Typical impurities in coke include compounds of calcium, iron, sulfur, titanium, vanadium, and boron. Most of these compounds boil away by the graphitization point, but certain metal oxides and carbides remain, as listed in Table~\ref{tb:impurities}. Boron has a deleterious thermal neutron capture cross section of 767~b; the other minerals capture about a hundred times less but are abundant enough to be undesirable in nuclear purposes as well. The simplest solution is to raise the peak furnace temperature to about 3,000$^\circ$C to exceed their boiling points; Table~\ref{tb:impurities} shows the remaining impurity levels after this ``thermal purification.'' 

The interactions of boron with carbon deserve a particular mention. Boron exists in graphite mostly as very hard, inert boron carbide (B$_4$C). It was thought in the 1940s that boron carbide remained when graphite is ashed. However, boron carbide melts at about the same temperature carbon graphitizes. So, when graphite is made, some of the decomposing boron carbide releases individual borons that get trapped within budding graphene crystals. When the graphite gets ashed, these ``substitutional borons'' oxidize and dissipate as boron monoxide (BO) or boric oxide (B$_2$O$_3$). Such boron transformation and loss mechanisms was not fully known until the 1950s~\cite{lowell67}.

% At this thermodynamic overlap, borons can get ``substituted'' for carbons as the graphite crystallizes; this mechanism was not known until the 1950s~\cite{lowell67}. Borons get freed again when the graphite is ashed, whereupon they oxidize into boron monoxide (BO) or boric oxide (B$_2$O$_3$) powders, and get carried away by the carbon monoxide (CO) or dioxide (CO$_2$) gases also produced. In other words, during the war, boron was thought to remain in ash as a carbide, but during graphitization, the carbide transforms into substitutional borons that, after ashing, get oxidized and lost by gas. % While Leo Szilard and Wilhelm Hanle guessed these oxidation mechanisms during the war, the fact that such substitutional borons could form was not known until the 1950s.

Because thermal purification cannot remove boron, it is important to identify low-boron cokes. Table~\ref{tb:graphites} lists the thermal capture cross sections and boron content for different graphites. All American graphites listed were made of petcoke, but Speer and AGOT were both made with a particularly pure Kendall petcoke from Pennsylvania. AGOT was thermally purified. Given pure carbon's capture is 3.2~mb---as we will calculate later---then 2.3~mb is the capture by ash in AGX graphite, 1.3~mb in Speer, and 0.9~mb in AGOT. Using the Kendall feed reduced capture by 1.0~mb (AGX vs.~Speer), and thermal purification cut it by another 0.4~mb (Speer vs.~AGOT). In other words, achieving AGOT nuclear-grade graphite was about two parts finding a low-boron coke and only one part purifying away metal oxides. 

% Roughly speaking, petcoke graphites have density 1.7 g/cm$^3$ and boron content 0.5~ppm; pitch coke graphites have density about 1.6 g/cm$^3$ and a few to tens of ppm boron~\cite{macpherson55}. This makes coal-derived cokes less desirable for nuclear purposes. %, but among them pitch coke is the most palatable. 
% These properties do vary geologically; not all petcokes necessarily suffice for nuclear-grade graphite either. 

\subsection{Siemens Electrographite}
On the German side, Siemens electrographite (SEG) from Siemens Plania Works was the ``the purest carbon that is manufactured in large quantities.'' It had 0.092\% ash and about 7~ppm by atom boron and 0.1~ppm by atom cadmium; we observe from Table~\ref{tb:graphites} that its impurities were higher than any graphite considered for Fermi's experiments. This certainly had to do with SEG's feed coke, but definitive internal data is scarce since Siemens Plania's scientific and head offices in Berlin-Lichtenberg fell under Soviet occupation. Their furnaces at Meitingen in the American zone were undamaged, though, and were surveyed by Anglo-American officers in late 1945~\cite{fiat397,fiat115,bios337,bios338}. 

% Germany depended on American crude oil imports for their electrographite until these stopped in January 1940, after which they turned to inferior substitutes containing more mineral impurities, like metallurgical coke and tar pitch coke from the rich coalfields in the Ruhr~\cite{scb40,fiat397}.  

Reportedly, Siemens Plania produced the highest quality and quantity of electrographite in Germany, using imported American petcoke until these ceased in January 1940~\cite{fiat397,scb40}. During the war, all electrographites had been rationalized, at least on paper, to just one grade for most industrial uses~\cite{fiat115}. As Germany's situation deteriorated though, SEG feeds not only worsened but increasingly varied, resorting to all sorts of coal-derived cokes~\cite{fiat397,bios337}. SEG was initially heated to 2,500$^\circ$C in the 1930s but reached 3,000$^\circ$C by 1945, undoubtedly to mitigate the rising impurities in poorer cokes, like the compounds listed in Table~\ref{tb:impurities}~\cite{bios337,pirani30}. This thermal purification may have been what a Siemens manager imagined when he claimed to Heisenberg in 1947 that a small quantity of purer SEG could have been supplied for the Uranverein in 1940 (\S\ref{tr:mueller} \cite{Mueller}). We exposit SEG's manufacture here because we are unaware of any prior discussion on this, yet it adds vital context to the nuclear properties of SEG, especially that SEG was eventually thermally purified like AGOT, albeit due to very different circumstances and feedstocks.

\begin{table}[t]\centering 
\caption{Representative impurities in unpurified and thermally purified petroleum coke-based graphites, their thermal capture cross sections (barns), equivalent boron content (EBC, ppm by mass), usual form in graphite, and boiling points~($^\circ$C)~\cite{glass22}. Impurity levels are from Ref.~\cite{eatherly62} and converted to EBC per Ref.~\cite{astm}. The 0.488 ppm EBC corresponds to the 0.5 ppm reported for AGOT in Table~\ref{tb:graphites}.}% Data from Refs.~\cite{eatherly62,BNL1,astm,glass22}.} 
\label{tb:impurities}
\begin{tabular}{l|r|cc|cc|cr}
\toprule
 & & \multicolumn{2}{c|}{Impurity level, ppm} & \multicolumn{2}{c|}{EBC, ppm}  & & \\
\midrule
Impurity & $\sigma_a^\text{th}$, b & Unpurified & Thermal pur. & Unpurified & Thermal pur. & Usual form & BP, $^\circ$C \\
\midrule
Calcium  & 0.4 & 320\phantom{.0} & 147\phantom{.0}                     & 0.049 & 0.022 & CaO & 2,850 \\
Iron     & 2.6   & 310\phantom{.0} & \phantom{0}10\phantom{.0}         & 0.201 & 0.006 & Fe$_2$O$_3$ & 2,623 \\
Sulfur   & 0.5 & 175\phantom{.0} & \phantom{0}19\phantom{.0}           & 0.040 & 0.004 & FeS & None \\
Titanium & 6.4 & \phantom{0}34\phantom{.0} & \phantom{0}11\phantom{.0} & 0.061 & 0.020 & TiO$_2$ & 2,972 \\
Vanadium & 5.0 & \phantom{0}30\phantom{.0} & \phantom{0}25\phantom{.0} & 0.042 & 0.035 & VC & c.~3,900 \\
Boron    & 767\phantom{.0} & \phantom{00}0.5  & \phantom{00}0.4        & 0.500 & 0.400 & B$_4$C & >~3,500 \\
\midrule
 & & & \multicolumn{1}{r|}{Total} & 0.893 & 0.488 & \\
\bottomrule
\end{tabular}
\end{table}

\begin{table}[] \centering
\caption{The ash content (in percent mass), thermal neutron capture cross section $\sigma_a$ (in millibarns), and equivalent boron content (EBC, ppm by mass) of German and American graphites. The original wartime values are taken from the corresponding references, which we corrected using modern cross sections for carbon, i.e., $\sigma_a^\text{C}=$~3.15~mb.}\label{tb:graphites}% 
\begin{tabular}{lllc|cc|cc|r}
\toprule
       &      &         &  & \multicolumn{2}{c|}{$\sigma_a$, mb} & \multicolumn{2}{c|}{EBC, ppm} & \\
\midrule
Author & Year & Sample  & Ash \% & Original & Corrected & Original & Corrected & Ref.\\ 
\midrule      
Bothe/Hanle & 1941/2 & Siemens & 0.092 & 9.2 & 7.7 &6.6&6.6& \cite{g71bothejensen,g153hanle} \\
Bothe       & 1944   & Siemens & 0.092 & 8.0 & 7.7 &6.6&6.6& \cite{bothe44} \\
% This paper  & 2024   & Siemens &  ---  & --- & 7.7 &---&---& \\
\midrule
Fermi et al. & 1942 & AGS/AGX  & 0.075 & 6.7 & 5.5 & 2.1 & 3.4 & \cite{nbs40-01-05,cp257}\\ % 6.68
Fermi et al. & 1942 & DCC      & 0.053 & 6.4 & 5.2 & 1.8 & 3.0 & \cite{nbs40-01-05,cp257}\\ % 6.38
Fermi et al. & 1942 & Speer    & 0.045 & 5.5 & 4.5 & 0.7 & 1.9 & \cite{ncc40-12-23,cp257}\\ % 5.49
Fermi et al. & 1942 & AGOT     & 0.030 & 5.0 & 4.1 & 0.5 & 1.4 & \cite{fermi52}\\ % 5.05
\midrule
% Lamarsh & 1966 & Reactor graphite & 1.70 & 0.022 &  4.2 & --- & --- & \cite{stansberry99,Lamarsh}\\
This paper & 2025 & Carbon  & --- & --- &  3.2 & --- & --- & \cite{BNL1}\\ 
\bottomrule
\end{tabular}
% \footnotetext{Source: This is an example of table footnote. This is an example of table footnote.}
\end{table}

\begin{figure*}[h!]

\end{figure*}

\section{Interpreting American and German Nuclear Data}\label{s:bjrepro}
During the war, there were two methods to measure impurities in graphite. One was by spectroscopy, which could identify individual elements in the sample, but the boron contents involved were near the detection limit of the era's best equipment~(\S\ref{tr:g153hanle}~\cite{g153hanle}). The other was by experimentally diffusing neutrons through the sample and measuring how far they survived (i.e., the diffusion length). This required knowing carbon cross sections accurately to subtract any effects by carbon itself. Furthermore, this method could not \textit{distinguish} impurities, so they were quantified in bulk as ``equivalent boron content'' (EBC), i.e., the ppm of boron that would cause the same absorption.

\subsection{Manhattan Project's Data}\label{s:2-nucleardata}
From 1942 onwards, the Manhattan Project used the diffusion method and listed graphite impurities as EBC. All EBCs were calculated relative to a carbon capture cross section $\sigma_a^\text{C}$ of 4.9~mb and scattering cross section $\sigma_s^\text{C}$ of 4.1~b. For example, for AGX graphite, they measured the diffusion length $L$ and derived its capture cross section $\sigma_a^\text{AGX}$ by:
\begin{equation}
    \sigma_a^\text{AGX} = \frac{1}{3 N^2 \sigma_s^\text{C} L^2} \approx \frac{12.8\cdot 10^{-24}}{L^2} = 6.7\text{ mb}, \label{eq:1}
\end{equation}
for atom density $N$~=~0.08 at/b-cm. Then, they got $y$ EBC by mass using:
\begin{align}
    \sigma_a^\text{AGX} &= x \sigma^\text{B}_a + (1-x) \sigma^\text{C}_a,\qquad  y = x \cdot a^\text{B} / a^\text{C} = 2.1\text{ ppm EBC}, \label{eq:2}
\end{align}
for $\sigma_a^\text{B}$~=~770~b, $x$ EBC by atom, and atomic masses $a$ for boron and carbon. 

However, their carbon cross sections are incorrect, chiefly that of scattering. Carbon's thermal scattering cross section known at that time was 4.1~b or 4.8~b measured at Columbia and Cambridge, respectively~\cite{dunning35,goldhaber37,weizsacker}. Fermi, Heisenberg, and Bothe used the former value in their calculations (the Germans rounded to 4~b). Contra that, a modern Monte Carlo simulation of neutron diffusion in graphite yields 4.93~b for scattering and 3.15~mb for capture. 
The one-fifth difference in scattering values causes roughly a 20\% error in all carbon capture cross sections measured by American and German physicists alike. Thus, to re-appraise the impurities in wartime graphites, Table~\ref{tb:graphites} lists their corrected capture and EBC by subbing the modern data into Eqs.(\ref{eq:1}) and (\ref{eq:2}). Siemens electrographite's impurities were measured both by diffusion and spectroscopy, so there is no need for EBC correction.

\subsection{Bothe's Experiment}\label{s:3-theory}
Heisenberg had predicted in 1939 that carbon's capture cross section should be 3~mb, which Bothe sought to verify with sufficiently pure samples of graphite. Walther Bothe and Peter Jensen's January 1941 experiment also used the diffusion method, diffusing neutrons from a radon-beryllium source through a ``tightly packed'' sphere of SEG bricks. 

By measuring a diffusion length $L$ of 36~cm, Bothe determined the thermal capture of the SEG to be:
\begin{equation}
    \sigma_a^\text{SEG,Bothe} =  \frac{1}{3 N^2 \sigma_s^\text{C} L^2} \approx 9.2\text{ mb}. \label{eq:xs_a^SEGBothe}
\end{equation}
Then, to differentiate the capture by carbon from that by impurities, Bothe incinerated the SEG and measured the ash to capture about 1.7~mb. Subtracting, the carbon capture cross section came out to 7.5~mb.  

To get Heisenberg's predicted 3~mb, the ash should have captured about three times more neutrons. But, in early 1940, Bothe had been informed by Hermann Sch\"uler at KWI for Physics that based on spectroscopic measurements of graphite impurities, ``calcium was the most prevalent; boron could not be detected'' (\S\ref{tr:g12bothe}~\cite{g12bothe}). It is unclear what sample Sch\"uler analyzed, but Bothe seemingly accepts and quotes this explanation for SEG's ash. All of the data at hand pointed to Heisenberg being wrong. Thus Bothe concluded, ``carbon even if it is produced with the best-known technical processes... can hardly be considered as a braking substance.''

It was only several months later that Wilhelm Hanle, at the University of G\"ottingen, measured SEG to have about 7~ppm boron and 0.1~ppm cadmium by atom, or 6.6~ppm EBC by mass (\S\ref{tr:g85hanle} \cite{g85hanle}; \S\ref{tr:g153hanle} \cite{g153hanle}). He noted these quantities were at the limit of detection for spectroscopy, although there was no other way, e.g., chemically, to detect them. Bothe apparently never directly learned of Hanle's results, but he commented on the possibility of ``a little boron'' in his 1944 journal article of this experiment. Importantly, there, he also used 4.5~b instead of 4~b as the carbon scattering cross section, lowering his capture values to 8.0~mb for SEG and 6.4~mb for pure carbon~\cite{reed-bothe}. 

\begin{comment}
\begin{figure}[]
    \centering
    \begin{subfigure}{.4\textwidth}\centering
    \includegraphics[width=\linewidth]{images/bothe.pdf}\caption{The graphite sphere was wrapped in cadmium foil and held by a rubber jacket, standing in a tank with at least 10~cm water on all sides.}
    \end{subfigure}\label{fig:sketch}
    \quad
    \begin{subfigure}{.4\textwidth}\centering
    \includegraphics[width=\linewidth]{images/flux-log-edit.pdf}\caption{Lin-log plots of neutron energy distribution in the graphite with and without the cadmium layer; their difference would isolate just thermal neutrons.}\label{fig:flux}
    \end{subfigure}
    \caption{Bothe \& Jensen's 1941 experiment.}\label{fig:experiment}
\end{figure}   
\end{comment}

\subsection{Correcting Bothe's Results with MCNP}\label{s:3-measure}
Now, we can reproduce this experiment using modern-standard Monte Carlo N-Particle (MCNP) code. A quick test in MCNP shows the carbon cross sections for the energy distribution of neutrons in Bothe's apparatus should be $\sigma_s^\text{C,MCNP}$ = 4.93~b for scattering and $\sigma_a^\text{C,MCNP}$ = 3.15~mb for capture. Suppose we take Bothe's raw data for granted and we solve for $L$ to a greater precision: 35.4338~cm. Using MCNP-calculated cross sections, the capture in SEG is now: 
\begin{equation}
    \sigma_a^\text{SEG} =  \frac{1}{3 N^2 \sigma_s^\text{C,MCNP} L^2} = 7.6765 \text{ mb}, \label{eq:4} % 7.473110 with l = 36 cm 
\end{equation}
% As mentioned in \S , using 4.0~b for scattering added roughly a 20\% bias to his calculations.  % Compared to Bothe's $\sigma_a^\text{SEG,Bothe}$ = 9.2~mb obtained 
roughly a 20\% change from Bothe's original value of 9.2~mb. Next, we examine whether Hanle's impurity measurements were accurate. In MCNP, we modeled SEG by drawing a sphere of pure carbon with density 1.68~g/cm$^3$ and then impregnated it with 7~ppm by atom boron and 0.1~ppm by atom cadmium. In that case, the capture in the SEG is:
\begin{equation}
    \sigma_a^\text{SEG,Hanle} = 7.7247 \pm 0.0001 \text{ mb}, \label{eq:5}
\end{equation}
which is a 0.6\% error from our above value. This ascertains Hanle's measurements to have been reasonably correct for SEG. We also double-checked by making sure subtracting the impurities retrieves $\sigma_a^\text{C,MCNP}$, like in Eq.(\ref{eq:2}). Hence, the more likely thermal capture in SEG should be about 7.7~mb and not 9.2~mb ($\sigma_a^\text{SEG,Bothe}$). 

\subsection{What Mistakes Did Bothe Make?}\label{s:3-mistakes}

We can now assess what mistakes Bothe made or did not make in this experiment. 
In our opinion, the most important and avoidable mistake was copying the incorrect scattering cross section of 4~b assumed by Heisenberg. Hypothetically, had Bothe used the 4.5~b value he did in 1944 (from averaging the Columbia and Cambridge data) and learned of Hanle’s findings, he could have calculated a carbon thermal capture of 5.0~mb, almost identical to the 4.9~mb used by the Manhattan Project. % using an identical procedure as ours, although he would have still used flawed boron and cadmium data, as explained in Appendix~\ref{s:2-error}. % This value would align with the 4.9~mb obtained by Manhattan Project physicists in 1942~\cite{cp257}. 

On the other hand, air gaps, which Heisenberg suggested were Bothe's mistake, were negligible. Heisenberg could have referred to macroscopic gaps between blocks or microscopic gaps within each one. In the former case, if there was just 1\% air, the capture in SEG would have noticeably ballooned from 7.7~mb to 17.7~mb, but we see good agreement between Eqs.(\ref{eq:4}) and (\ref{eq:5}). Plus, Bothe mentions the apparatus was ``tightly packed''; when he measured a pitted, inhomogeneous graphite earlier, he dutifully reported much higher errors~\cite{g12bothe}. Regarding microscopic gaps, all manmade graphites are about 25\% porous, but this is clearly reflected in SEG's density in Eq.(\ref{eq:xs_a^SEGBothe}). For these reasons, air gaps do not seem to have palpably skewed any results. 

Bothe does dismiss the possibility of boron impurities, but it is unfair to blame him. Not only had Sch\"uler's spectroscopy found no boron in electrographite, he fairly reasoned that any impurities which survive graphitization at 2,400$^\circ$C would not be boiled away when ashing at 700$^\circ$C. The mechanism by which substitutional boron forms and escapes during ashing was only discovered in the 1950s~\cite{lowell67,hennig65}. For the two boron compounds produced during ashing, the discovery of boron monoxide had only been announced by IG Farben in October 1940, just three months before the experiment, and the phase change temperatures of boric oxide were unknown~\cite{zintl40}.

\section{Comparative History of Graphite Experiments}\label{s:4}
% The historical record of the Uranverein's investigation of graphite as a suitable reactor moderator can be traced through document-based nuclear archaeology of primary sources. 

It would thus be worthwhile to reconstruct what German and American physicists knew in the 1940s to better understand how their actions were informed, especially since these experiments were conducted by Nobel-caliber minds like Heisenberg (1932 winner), Fermi (1938), and Bothe (1954). This is underscored by remarkable parallels in both programs' graphite experiments. % In fact, a detailed comparative study will show how Fermi also believed air gaps were the sole impurities in his first graphite measurement and could not explain why he lost most of the expected boron in his ash. 

\subsection{Early German Experiments, 1940}\label{s:4-40}
By the invasion of Poland on September 1, 1939, the Army Ordnance Office (Heereswaffenamt) managed the work of nuclear physicists informally called the Uranverein, split across various institutions in Germany. Scientists submitted their findings to the Ordnance Office, which then distributed the reports on a ``need-to-know'' basis. After the war, the American ALSOS mission saved 394 of these papers~\cite{GList}, enabling document-based nuclear archaeology of their investigations with graphite, of which we translate the most relevant in the Appendix.

The Germans' nuclear knowledge at this intellectual junction is captured in a two-part report by Werner Heisenberg for the Army Ordnance Office, split into Part I dated December 9, 1939 and Part II dated February 29, 1940 (\S\ref{tr:g39heisenberg}~\cite{okw1}; \S\ref{tr:g40heisenberg}~\cite{okw2}). In them, Heisenberg discussed trade-offs between reactor materials, favoring enriched uranium with light water moderation as the most reliable design. However, the prohibitive cost of enrichment limited options to heavy water or carbon, which could moderate natural uranium. As no sufficiently pure sample of carbon had been isolated for accurate measurements, Heisenberg theorized carbon's thermal capture cross section to be 3.0~mb, assuming 4~b for scattering, a critical error as discussed in \S\ref{s:3-mistakes}. The data on carbon, though, was so uncertain that Heisenberg worried heavy water would be the only viable option.

In 1940, the Uranverein began testing light water, heavy water, and carbon moderation using alternating uranium and moderator plates. Karl-Heinz H\"ocker calculated that carbon moderation yielded ten times less neutron multiplication than heavy water for the same mass of uranium---a finding consistent with the critical sizes of Chicago Piles~1 and~3. Addressing Heisenberg's uncertainties about carbon, H\"ocker predicted that a capture cross section of 7.8~mb was the maximum at which neutron multiplication could still occur (\S\ref{tr:g41hocker}~\cite{g41hocker}; \S\ref{tr:g42hocker}~\cite{g42hocker}).

In March 1940, Georg Joos and Wilhelm Hanle at the University of G\"ottingen analyzed boron and cadmium levels in commercial carbon materials, like graphites, medicinal charcoals, and carbohydrates (\S\ref{tr:g46joos}~\cite{g46joos}, \S\ref{tr:g35hanle}~\cite{g35hanle}). Apparently, electrically heating graphites with ``300 amps for a short time'' failed to remove any boron, as it persisted as ``low-volatile boron carbide.'' This ruled out thermally purifying out boron in carbon. The G\"ottingen physicists had not tested electrographite though, but Sch\"uler at KWIP had found no boron in a sample of one. As long as boron was absent, heat could purify all the other mineral impurities. Walther Bothe at KWImF in Heidelberg, having read the reports from both institutions, thus remained optimistic in the viability of electrographite (\S\ref{tr:g12bothe}~\cite{g12bothe}).

\subsection{The German Army Ordnance Decision, 1941--1942}\label{s:4-4142}
Upon these scientific bases, Bothe and Jensen tested electrographite from Siemens Plania in January 1941 (\S\ref{tr:g71bothejensen} \cite{g71bothejensen}). As we showed in \S\ref{s:3-measure}, an underestimated carbon scattering cross section led them to overestimate capture in the graphite to 9.2~mb and that in carbon to be ``7.5~$\pm$~1~mb,'' straddling the maximum allowance of 7.8~mb calculated by H\"ocker. Still, Bothe concluded that  graphite was unsuitable as a reactor moderator without uranium enrichment. Bothe's dismissal of any boron impurities, partly influenced by Sch\"uler's results, were contradicted in April by Wilhelm Hanle (\S\ref{tr:g85hanle}~\cite{g85hanle}). By March 1942, Hanle also found a sugar from Pfeifer \& Langen with practically no boron contamination that could be a pure carbon reference, with the caveat that: ``This would probably only be worthwhile if carbon is really still considered for the [moderation] problem. In any case, part of the neutron capture in the graphite used by Bothe can be explained by boron contamination''~\cite{g153hanle}. 

% Bothe's dismissal of any boron impurities, partly influenced by Sch\"uler's results, were contradicted in April by Wilhelm Hanle (\S\ref{tr:g85hanle}~\cite{g85hanle}). The problem, Hanle noted, was that boron could not be clearly distinguished from carbon because there was no known pure carbon reference, and the boron quantities involved were so small that it was at the limit of spectroscopy's detection, but there was no other way was sensitive enough. Instead, he measured the boron content by adding known amounts of it until its spectral lines doubled, though this method was imprecise, varying within samples and detecting very trace, but non-zero, boron. By March 1942, Hanle found a sugar from Pfeifer \& Langen with practically no boron contamination that could be a pure carbon reference, with the caveat that: ``This would probably only be worthwhile if carbon is really still considered for the [moderation] problem. In any case, part of the neutron capture in the graphite used by Bothe can be explained by boron contamination''~\cite{g153hanle}. 

But Hanle was too late. It is important to briefly unpack the national context in Germany at this time. Because of a vertiginous rearmament program without any foreign currency for necessary imports, every sector had been operating under austerity measures since at least 1937~\cite{tooze-ch8}. By 1941, steel production was only about half of what the Wehrmacht demanded, and coal and oil were in deficits of tens of millions of tons, making scarce their coke byproducts~\cite{tooze-ch10}. Coke in turn was a vital ingredient in steel and aluminum production, so much so that Hitler announced to his advisors with brutal clarity: ``if, due to the shortage of coke the output of the steel industry cannot be raised as planned, then the war is lost''~\cite{tooze-ch15}. Deeply cognizant of Germany's situation, Hitler had wagered all military-industrial planning on quick ``lightning wars,'' but the calculus lurched for the worse in December 1941: the Japanese attacked Pearl Harbor and fresh Soviet reserves decimated Army Group Center just 100~km from Moscow. Now faced with a drawn-out war of attrition amid staggering losses, Army Ordnance declared that the Uranverein's work could only be justified ``if it is certain that an application will be achieved in the foreseeable future''~\cite{walker24-hwa}. 

This precipitated a formal Army review of the Uranverein's progress, entitled \textit{Energy Production from Uranium} and dated February 28, 1942 (\S\ref{tr:energie} \cite{hwa42}). This report concluded carbon was unsuitable for reactor moderation due to its capture cross section. The Army recognized that this carbon sample turned out to have small boron impurities, but it was ``practically impossible to produce carbon with a higher degree of purity''---probably referring to Joos's purification attempt and Bothe's comment that SEG was ``probably the purest carbon that is manufactured in large quantities.'' Because this work would not be immediately decisive for the war, the work was transferred to the civilian Ministry of Education, and the physicists redoubled their focus into heavy water experiments.

\subsection{American Experiments, 1939--1942}\label{s:4-americanexp}
Across the Atlantic, Leo Szilard and Enrico Fermi at Columbia University were also exploring the feasibility of graphite moderation. In July of 1939, three months after a chain-reacting ``uranium machine'' was proposed by Hanle and Joos in Germany, Szilard had already begun inquiring various companies for their purest graphite, of which samples from the National Carbon, Speer Carbon, and U.S. Graphite companies were identified. From New York, Szilard wrote to Fermi, then visiting the University of Michigan: ``Pending reliable information about carbon we ought perhaps to consider heavy water as the favorite... % july 5
A chain reaction with carbon is so much more convenient [than] heavy water... that we must know in the shortest possible time whether we can make it go''~\cite{szilard39-07-08}. % july 8

% In a nearly identical experiment, the Americans would face the same problem of impurities lost in ashing, just two weeks ahead of Bothe and Jensen in January 1941. The critical divergence occurred from Szilard and Norman Hilberry's continued persistence in wrangling the industry to actualize graphite purification. 

% As the Manhattan Project would only be convened in September of 1942,\footnote{The U.S., while involved in the Atlantic, would not formally enter World War II until the attack on Pearl Harbor on December 7, 1941, after which the Manhattan Project was convened on August 13, 1942.} much of this work was still conducted by Enrico Fermi's group at Pupin Laboratories of Columbia University, with Leo Szilard facilitating much of the materiel logistics with meager funds.

Szilard first wanted 4 short tons of the purest graphite to figure out the elemental carbon capture cross section, % followed by a scaled-up moderation experiment with 40 tons, 
but apparently, no single customer in America had ever ordered graphite in such great quantities before \cite{ncc39-07-06}. In 1939, from National Carbon, Szilard obtained electrode-grade 2301 graphite powder with as little as 0.02\% ash, but only their AGS graphite plates with ash content up to 0.1\% could be provided on the order of several tons. From U.S. Graphite, he received DCC Graphitar with reported ash content of 0.07\% and being vanadium-free. All of these graphites were made from petroleum coke~\cite{szilard40-12-11,szilard41-02-20}; by January 1940, the National Bureau of Standards measured the AGS and DCC to have 0.075\% and 0.053\% ash by mass, respectively, after proper dehydration~\cite{nbs40-01-05}. % He did not purchase any Speer electrographite as they could only report their ash content to about 0.5\%. 

In April 1940 at Columbia, H. L. Anderson and Fermi determined the thermal capture in pure carbon to be 3.0~mb using 4 short tons of either AGS or DCC~\cite{a21anderson}. DCC graphite was likely used, as it had the lower ash content measured by the NBS, and several tons had been sent by USG in March~\cite{usg40-03-01}. This aligns with Fermi's recollection of unwrapping newly-arrived graphite just prior to the experiment and M. D. Whitaker's account that USG's graphite was used in the ``early'' Columbia experiment~\cite{c133whitaker}. Anderson and Fermi only considered water vapor or nitrogen (air) gaps in the atomic sheets of graphite to be possible impurities, so it is very likely Fermi got such an optimistic value from commercial graphite due to the errors in his experimental design described in Appendix~\ref{s:2-error}. A repetition of this measurement at the Met Lab in 1942 yielded a slightly more reasonable result of 4.9~mb~\cite{cp257}.  % , as we will discuss in Section \ref{s:3-norm} Still, this is an exceedingly low value, as the modern accepted value for pure carbon is 3.8 mb and for reactor-grade graphite 4.2 mb~\cite{Lamarsh}. Anderson and Fermi's setup was quite standard, as described in \S\ref{s:2-error} and almost identical to same as Bothe and Jensen's. Furthermore, 

Szilard must have also purchased some graphite from the Speer Carbon Company by late 1940, as he sent a few pounds of it to NBS to be ashed and to National Carbon to be spectroscopically measured for impurities~\cite{szilard40-12-18,nbs41-01-27}. In December 1940, National Carbon physicist Herbert MacPherson\footnote{Unbeknownst to Szilard, MacPherson had completed physics doctorate at Berkeley in 1936 and been up-to-date on nuclear moderation and boron's high neutron capture, so the role of this graphite ``aroused speculation within National Carbon''~\cite{weinberg94}.} reported 0.78~ppm boron content by weight, with ash content between 0.045\% to 0.06\% \cite{ncc40-12-23}. Both MacPherson and Szilard surmised ``no chance'' of cadmium impurities for being too volatile to remain after graphitization, and followup measurements by MacPherson found none~\cite{szilard40-12-28, ncc40-12-30, ncc41-01-08}. % MacPherson recognized from his experience in high-purity carbon arcs for film projectors the importance of trace boron impurities, taking special care in this measurement~\cite{weinberg94}. : ``My first knowledge of the possible use of graphite to aid in sustaining a nuclear chain reaction came from reading [an article] in September 1939. This

In the first few days of January 1941, Fermi personally measured capture in Speer graphite's ash to verify MacPherson's boron measurements. Much to his surprise, Fermi observed less than half of the capture reported by MacPherson (0.53~mb vs. 1.22~mb)~\cite{szilard41-01-09}. Just two weeks later, Bothe and Jensen conducted the exact same experiment with similar results. The missing boron stumped the Columbia team. Szilard proposed to MacPherson on January 17: 

\begin{displayquote}
    I am rather puzzled by the discrepancy between your spectroscopic measurement of boron and Fermi's absorption measurement of the ash. From the known boiling point of \ce{B_2O_3}, which is only listed as ``greater than 1500$^\circ$,'' one cannot at all exclude the possibility that when graphite is burned... a considerable fraction of the \ce{B_2O_3} is carried away as vapor by the carbon dioxide which is produced. In [these] circumstances we fear that perhaps only one-third of the boron content of the graphite is present in the ash. \cite{szilard41-01-17}
\end{displayquote}

Szilard's hypothesis was remarkably similar to that of Hanle, yet MacPherson's response echoed Bothe's reasoning:

\begin{displayquote}
    In regard to the possibility of loss of boron by evaporation in the ashing process, it is my opinion, for theoretical reasons, that such a loss would be small. I therefore do not think it worth while for us to undertake any other work on the determination of boron in your graphite. \cite{ncc41-01-22} % If much boron were to be lost by evaporation... one might expect to find a significant difference in the amount of boron found in [other experiments].
\end{displayquote}

Undeterred, Szilard continued to pepper both U.S. Graphite and National Carbon staff with ideas to detect and separate boron from petcoke feed, eventually incurring a steely response from the latter that to produce boron-free graphite would ``require a time consuming research of a magnitude that we do not consider desirable to undertake'' \cite{ncc41-01-31,ncc41-02-26}. Szilard wrote in a funding request that February, ``boron appears to be present in the ash and the amount actually contained in the graphite may be more as some of the process is lost in ashing.... Experiments may show that much purer graphite than that available at the present time is important.'' He commented that the other ongoing efforts with berllyium or heavy water moderation seemed more promising.

\subsection{Birth of Nuclear Graphite, 1942} \label{4-nucleargraphite}

In the winter of 1941--42, physicists Norm Hilberry from New York University and Samuel K. Allison from the Met Lab traveled to the Speer company in Pennsylvania. In an effort to improve the conductivity of their electrodes, Speer had discovered that petcoke from the nearby Kendall Oil Company was exceptionally crystalline and had unusually little mineral, boron, or sulfur impurities \cite{hilberry65}. Compared to cokes from Texas and Louisiana, Kendall petcoke was indeed ``the purest cokes available to the industry,'' but the physicists had a few improvements to suggest~\cite{eatherly81}. Namely, the metcoke and pitch lining, to add resistivity and density, introduced impurities and were removed, though this reduced graphite yield from petcoke by half---deemed acceptable as long as the purity problem was solved.

Hilberry and Allison arranged for a batch to be sent to Fermi, now at Chicago, where in May 1942 a pile with the new Speer graphite achieved a tantalizing $k$ of 0.995---this was the lucky break the physicists needed, as it now became evident that a larger pile, with just a little purer graphite, could achieve criticality~\cite{allison62,c133,reed-piles}. Speer was a small operation, and removing the linings had cut into yield rates, but perhaps National Carbon---the largest of the firms---could be leveraged to pump out 250~tons of low-boron graphite for a large-scale experiment by the year's end. % Now, they had a promising game plan.

During a meeting between Szilard, Hilberry, and National Carbon's president at the New York City headquarters, Hilberry enunciated that moderation experiments were so important he could obtain Priority X to get the graphite they needed. The highest of all war production ratings, Priority X orders demanded precedence before all others, at the fastest possible production and delivery timelines. The next morning, a liaison phoned Hilberry:

\begin{displayquote}
    My instructions [from Washington] are to do what you tell me to do and shut up. The highest priority that’s been applied in the graphite business to date has been a C. When you walked in there and said you would give [National Carbon] an A priority on the spot and if you had to have an X you’d have it for them the next day, you just stopped the entire graphite business. \cite{hilberry65}
\end{displayquote}

The War Production Board commandeered National Carbon's production capacity within the week, and scientists MacPherson, Victor Hamister, and Lauchlin Currie were assigned to ensure the graphite met Fermi's desired purity. Working closely with Hilberry, Allison, and the National Bureau of Standards, the furnaces' peak temperature was increased from 2,400$^\circ$C to about 3,000$^\circ$C and doubled in duration to sufficiently burn away impurities~\cite{macpherson55,Smyth}. New handling procedures, including special wrapping paper, dry machining, and laundering all the workers' clothes with non-boron soaps, were also implemented. Finally, a test pile with National Carbon's new ``AGOT'' nuclear-grade graphite in November verified it had achieved even better purity than Speer's in May~\cite{cp341}. When the last shipment of AGOT arrived on December 2, 1942, Fermi immediately added it to the pile under Stagg Field and achieved a critical reaction---the now-famous Chicago Pile~1~\cite{allison65}.

\subsection{Similarities and Differences in Experiments}

We can now collate direct parallels in German and American explorations of carbon moderation. Both programs shared a genesis that incomplete carbon data in prewar literature led to uncertainties of its feasibility in reactors. As Szilard told Fermi in 1939, ``lacking reliable information about carbon, we ought perhaps to consider heavy water as the favorite,'' Heisenberg wrote regarding carbon, ``the  data available so far are still too inaccurate to make a final decision.'' Heisenberg had theorized carbon's thermal capture cross section to be 3.0~mb; Fermi obtained the same value experimentally in April 1940, albeit almost certainly by mistake. 

Both programs spent 1940 surveying impurities in commercial carbon products, from which certain graphites were short-listed. In January 1941, Fermi and Bothe ashed their graphites within two weeks of each other, and both could not explain how they were short over half the expected impurities in the ash. While MacPherson and Bothe ruled out any boron loss during ashing, Szilard and Hanle pointed out possible vaporization mechanisms of boron oxides, and neither side knew the complete answer involved substitutional boron impurities in the graphene lattice. American manufacturers insisted there was no way to control boron in coke, while Joos attempted to heat away boron in graphite to no avail. By 1942, Germany's polycrises jeopardized their Army's interest in the whole nuclear program. 

At that point, American progress with graphite was not due to any breakthrough in theory but rather when a uniquely crystalline and pure petroleum coke was found in Pennsylvania. Using this, plus thermal purification and some process controls, the new AGOT graphite allowed Fermi to achieve a self-sustaining chain reaction. Most ironically, despite this achievement, neither program actually calculated the correct capture cross section for carbon. This was because everyone had anchored their data on an inaccurate carbon scattering cross section measured at Columbia in 1935.

\section{Conclusions}
After the war, Werner Heisenberg claimed that Walther Bothe's pessimistic neutron absorption measurement in graphite mistakenly steered the Uranverein toward suboptimal heavy water pile designs. Some secondary histories have further suggested that where Bothe failed, Fermi ostensibly measured the correct carbon capture cross section in April 1940 and that this solved the dilemma on whether to commit to graphite-moderated reactors~\cite{rhodes12}. This narrative is perhaps potent in the popular consciousness because it canonizes the first triumph of American, industrialized science in a close race where if not for a few mistakes, Germany too could have had a nuclear reactor. 

We do not believe that interpretation is supported by empirical evidence nor a document-based reconstruction of wartime nuclear knowledge. Accusations levied personally against Bothe, such as air gaps or neglecting boron in his apparatus, were found to be unsubstantiated or unfair. Abandoning graphite moderation was not caused by a scientific error but rather from a calculated assessment of materials shortages and wartime priorities. With that much boron in Siemens electrographite, no high-quality coke to make better alternatives, and coke otherwise in critically short supply for steelmaking, graphite reactors were a genuinely infeasible proposition for Germany in 1942. Even in America, commercial graphites offered such razor-thin margins for nuclear multiplication that Szilard and Fermi seriously considered beryllium or heavy water moderators instead, until the Kendall petcoke was discovered.

Not enough recognition has been given to the influence of the vast American petroleum industry in facilitating nuclear-grade graphite manufacture. Reframing what was thought the Uranverein's failure to a deliberate decision uncovers a much more profound interplay between construction expediencies versus material scarcities in early reactor experiments. This certainly contests any ex post facto prejudices that a Chicago Pile 1-type need have been the only path to a wartime reactor or implications that the Uranverein's heavy water efforts were any more of a ``mistake'' than graphite. Hypothetically, even if the Uranverein committed to a pile with Siemens electrographite, it would have to be impractically larger than CP-1, as a former student of Bothe showed in 1980~\cite{koester80}. Or, they would have needed to master chlorine gas purification to flush out the boron, which would not be invented until 1947~\cite{eatherly81}. Such counterfactuals would also need to be justified without fantastically reinventing the Reich's warplanning priorities in 1941--42. 

Pragmatically, our work also demonstrates how pernicious antiquated assumptions or data, such as outdated cross sections, may be in historical decisions, and why a careful document-based nuclear archaeology to reconstruct their state of knowledge is necessary to fairly interpret them. This applies not only to historical data but to that of foreign states, where we may need to be cognizant of outdated, different, or implicit nuclear conventions. In this spirit, we hope our translations of German primary sources in the Appendix below enable readers seek their own conclusions.

\section*{Acknowledgements}
Our thanks goes to Noah Schwartz (Twente) and Tomasz Mikurda (AGH Krakow) for consults on our translations; to Joshua Govus (Harlow) for helping us obtain some FIAT reports in British archives; to Aayan Thapa (Sherfield), Alex O'Neill (CSUF), Dan Ho, Jason Zhao (TAMU), Jet Situ (CMU), Eugene Ahn (Berkeley), Hannah Hui (UCSD), Francis Chen, Eric Lepowsky, Jihye Jeon, Raven Witherspoon, Chang Yu (Princeton), B. Cameron Reed (Alma), Mark Walker (Union), and two anonymous reviewers for reading through this paper and providing very useful feedback.

\section*{Declarations}
% Some journals require declarations to be submitted in a standardised format. Please check the Instructions for Authors of the journal to which you are submitting to see if you need to complete this section. If yes, your manuscript must contain the following sections under the heading `Declarations':
\begin{itemize}
%\item Funding. Not applicable.
%\item Conflict of interest/Competing interests (check journal-specific guidelines for which heading to use)
\item \textbf{Competing interests.} The authors declare no competing interests.
%\item Ethics approval and consent to participate. Not applicable.
%\item Consent for publication. Not applicable.
%\item Data availability. Not applicable.
%\item Materials availability. Not applicable.
%\item Code availability. Not applicable.
\item \textbf{Author contribution.} PP wrote the paper; PP and SH wrote the translations; TK initiated, reviewed, and supervised the project; PP, SH, and TK consented to publication.
\end{itemize}

\begin{appendices}
\section{Systemic Error in WWII-era Thermal Capture Measurements} \label{s:2-error}

Early thermal capture experiments, relying on incomplete data for cadmium capture, had systemic inaccuracies. % ; and 2) neutron energy distributions in these measurements were not standardized, causing \textit{imprecisions} in the values.
In 1935, Fermi devised a method to measure thermal cross sections by comparing reaction rates in a sample with and without a cadmium shield, mistakenly assuming cadmium only absorbed neutrons below 1 eV (thermal energies)~\cite{fermi36a}. Unshielded measurements detected all neutrons, while cadmium-shielded ones excluded thermal neutrons, so the difference indicated the thermal neutron cross section:
\begin{equation*}
    \sigma_\text{unshield}(\text{fast + interm. + thermal}) - \sigma_\text{shield}(\text{fast + interm.}) = \sigma_\text{diff}(\text{thermal}).
\end{equation*}
This approach became standard, even adopted by Bothe in 1941. However, modern data from Brookhaven reveal that cadmium appreciably absorbs neutrons higher than 1~eV. So, the cadmium-shielded measurements actually excluded thermal \textit{and} some intermediate energy neutrons, so the difference in the two measurements was more like:
\begin{equation*}
    \sigma_\text{unshield}(\text{fast + interm. + thermal}) - \sigma_\text{shield}(\text{fast + some interm.}) = \sigma_\text{diff}(\text{thermal + some interm.}).
\end{equation*}
In other words, all the neutrons cadmium was capturing, Fermi thought were thermal, when some of them were in fact intermediate neutrons, meaning he overestimated cadmium's thermal capture. Because the difference $\sigma_\text{diff}$ represented in reality a higher average energy of neutrons than he expected, by the $1/v$ relationship, this caused Fermi to \textit{underestimate} thermal capture in elements measured using this cadmium method, such as boron and carbon. A comparison of the 1930--40s vs.~modern data in Table~\ref{tb:bcd} shows how thermal capture in cadmium was overstated, causing an underestimation of capture in boron. 

% In April 1940, Fermi's group at Columbia measured a thermal capture cross section of 3.0~mb in supposedly nearly pure graphite, but Fermi recognized this was a gross underestimation while at the Metallurgical Laboratory in 1942; see his papers, Refs.~\cite{cp74,c190fermi}. Other Manhattan Project physicists then repeated the measurement with Fermi's graphite and obtained a thermal capture of 4.9~mb, which is more realistic given they thought carbon scattering was 4.1~b (modern: 4.9~b). % Meanwhile, the Uranverein appears to have overlooked this issue entirely.

We also note here that Brookhaven gives the thermal capture in carbon to be 3.8~mb and in boron to be 767~b, but we used 3.15~mb and 619~b in the paper. Effective cross sections during diffusion differ from reference data because thermal neutrons get captured as they moderate from 1~eV to 0.0253~eV, and Fermi's indium or Bothe's dysprosium foils counted the capture of all neutrons sub-1~eV, whereas Brookhaven's values are based on the capture of an ideal Maxwellian energy distribution of neutrons modal at 0.0253~eV. To represent realistic neutron energies, MCNP-simulated cross sections triumph for our purposes.

\begin{table}[h!]\centering
\caption{The reported thermal capture cross sections (in barns) for elemental boron and cadmium from pre-war and Uranverein sources, plus the 2018 reference value for thermal neutrons from Brookhaven.} \label{tb:bcd}
\begin{tabular}{l|llrrr}
\toprule
Era & Year & Source & B [b] & Cd [b] & Ref. \\ 
\midrule
\multirow{4}{*}{Pre-War} & 1935 & Amaldi \& Fermi & 3,000  & 10,000 & \cite{amaldi35}  \\
& 1935 & Dunning et al. & 360 & 3,300 & \cite{dunning35} \\
& 1936 & Fink & 540 & 2,900 & \cite{fink36} \\
& 1937 & Griffiths \& Szilard & --- & 4,500 & \cite{griffiths37} \\ 
\midrule
\multirow{4}{*}{Uranverein} & 1940 & Haxel \& Volz & --- & 3,670 & \cite{g37haxelvolz} \\
& 1940 & Reddemann \& Bomke & 390 & 2,640 & \cite{reddemannbomke} \\
& 1941 & von Droste & 545 & --- & \cite{g76droste} \\
& 1941 & Kremer & 508 & 3,480 & \cite{g104kremer} \\ 
\midrule 
\multirow{1}{*}{Manhattan} & 1942 & Fermi & 770 & --- & \cite{cp74} \\ 
\midrule 
Modern & 2018 & Brookhaven & 767 & 2,520 & \cite{BNL1} \\ 
\bottomrule
\end{tabular}
\end{table}

\section{Translations}
Headers denoted without an ``excerpt'' are full translations.

\subsection{Werner Heisenberg}
\subsubsection{Excerpt: ``The Possibility of Useful Energy Production from Uranium Fission---Part I,'' G-39 (December 6, 1939)}\label{tr:g39heisenberg}
[p.385] The following table [Table \ref{tb:g39}] provides a compilation of cross sections for capture ($\sigma_\text{r}$) and for elastic scattering with thermal neutrons ($\sigma_\text{th}$), with neutrons of 25 eV ($\sigma_\text{25}$), and with fast neutrons ($\sigma_\text{sch}$) in the elements H, D, C, O, and U. The values have been obtained partly from measurements and partly from theoretical estimates. Well-determined values are underlined (printed in boldface). For the other values, errors of 50\% or more are entirely possible. The cross sections are given in units of $10^{-24}$ cm$^2$.

\begin{table}[h!]
    \centering
    \caption{}\label{tb:g39}
    \begin{tabular}{lllllll}
        \toprule
         Substance & & H & D & C & O & U \\ 
        \midrule
         Capture cross section & $\sigma_\text{r}$ & \phantom{0}\textbf{0.3} & 0.003 & 0.003 & 0.003 & \phantom{0}\textbf{3.4} \\
         \midrule
          & $\sigma_\text{th}$ & \textbf{40} & 7 & 4 & 3.3 & 10 \\
         Scattering cross sections & $\sigma_\text{25}$ &  \textbf{13} & 3 & 3.5 & 3 & --- \\
          & $\sigma_\text{sch}$ & \phantom{0}2 & 2 & 2 & 2 & \phantom{0}6 \\
        \bottomrule
    \end{tabular}
\end{table}

[p.396] The fission processes in uranium discovered by Hahn and Strassmann can, based on the data available so far, also be used for large-scale energy production. The safest method for constructing a machine suitable for this purpose is the enrichment of the isotope uranium-235. The further the enrichment is carried out, the smaller the machine can be built. The enrichment of uranium-235 is the only method by which the volume of the machine can be reduced to less than 1 cubic meter... However, for energy production, normal uranium without enrichment of U-235 can also be used if uranium is combined with another substance that slows down the neutrons from uranium without absorbing them. Water is not suitable for this purpose. In contrast, according to the data available so far, heavy water and highly pure carbon fulfill this purpose.~\cite{okw1}

\subsubsection{Excerpt: ``The Possibility of Useful Energy Production from Uranium Fission---Part II,'' G-40 (February 29, 1940)}\label{tr:g40heisenberg}
[p.397] The following report contains a detailed elaboration of the considerations that were briefly outlined in my earlier report... Section II.4 includes a new calculation of the key data for the uranium machine, based on the new experimental results and more precise theoretical formulas. This section also corrects an error that occurred in Part I, Table 1, which Bothe brought to my attention. The results of Section 4 indicate that the conditions for the construction of the uranium machine as assessed in Part I may have been somewhat too favorable. In particular, it has become doubtful whether the uranium machine could be constructed using pure carbon. The experimental data available so far are still too inaccurate to make a final decision. Otherwise, not much changes in the information provided in Part I.

[p.418] If $\sigma_r$ is approximately $0.003\cdot10^{-24}$ cm$^2$ as assumed in Part I, then pure carbon is not sufficient for constructing the uranium machine. However, the conditions could improve with the use of uranium plates. Configurations where uranium layers alternate with layers of another substance will not be further addressed in this report, as von Weizs\"acker has conducted detailed calculations on this subject and will likely communicate them soon.~\cite{okw2}

\subsection{Georg Joos, ``The Production of Extremely Pure Carbon,'' G-46 (March 29, 1940)}\label{tr:g46joos}

To the Army Ordnance Office, Research Department 1:

The most promising way to achieve a chain reaction at the moment is neutron moderation using carbon. But this must be extremely pure. The most dangerous impurity is boron due to its enormous absorption cross section. A concentration of $10^{-6}$ is just about tolerable, since then the total cross section of the boron amounts to 10\% of the absorption cross section of carbon assumed by Heisenberg. It was stated by V. M. Goldschmidt in 1932 (G\"ott.~Nachr.~1932, p.~403) that it is hopeless to remove boron from natural coals. Since then, however, the purification process has been improved. In a method developed at the Mineralogical Institute at G\"ottingen, coal is exposed for a short time to a current load of 300 A. Usually, due to the high vapor pressure inside, an explosive shattering occurs. With this method, arc lamp carbons and rods of the purest Acheson graphite were treated in the high-current facility of the Mineralogical Institute. Unfortunately, the spectroscopic examination showed that even with this treatment, the boron, which apparently forms a low-volatile boron carbide, cannot be removed. So, carbon was also sampled from heating carbohydrates.

The spectral analysis was carried out by Prof.~Hanle using the boron doublet lines at 2497/8 [{\AA}]. Exposures were calibrated so that these lines become just visible at a concentration of $10^{-6}$. Coals in which no trace of the doublet can be seen on the plate even with long exposure are therefore acceptable. The results are as follows [Table \ref{tb:g46joos}]:

\begin{table}[h!] \centering
\caption{}\label{tb:g46joos}%
\begin{tabular}{lc}
\toprule
Type of coal & Atomic concentration of B \\
\midrule
Bitmuminous coal & $10^{-5}$ to $10^{-6}$  \\
Beechwood [hardwood] charcoal & $10^{-5}$  \\
Graphite from Acheson Company [imported, see~Ref.\cite{pirani30}] & $10^{-4}$  \\
`` '' heat-treated & $10^{-4}$  \\
Arc lamp carbon, heat-treated & $10^{-4}$  \\
Granulated activated charcoal, Merck & $10^{-5}$  \\
Powdered medicinal carbon (blood charcoal), Merck & $10^{-5}$  \\
\midrule
Carbon from rock candy    & $<10^{-6}$  \\
Carbon from table sugar    &  $<10^{-6}$ \\
Carbon from glucose    & $<10^{-6}$ \\
Carbon from potato starch & $<10^{-6}$ \\
\bottomrule
\end{tabular}
% \footnotetext{Source: This is an example of table footnote. This is an example of table footnote.}
\end{table}

Thus, only the last 4 types of carbon remain in consideration. Rock candy yields the purest carbon, as it originates from a crystallization process, but ordinary table sugar also meets the requirements. Glucose shows a large number of weak, as yet unidentified impurity lines. Carbon from potato starch yields an atomic concentration of $10^{-3}$ for phosphorus and potassium and $10^{-4}$ for calcium. To what extent these impurities are harmful cannot be decided based on the material available here about effective cross sections. The packing density of the very foamy carbon from sugar was brought to 1.0 [g/cm$^3$] by sintering. However, it can probably be brought to the value of 1.25 assumed by Heisenberg by sintering. Based on these results, it is proposed to have a few kilograms of carbon produced from sugar and starch in a suitable chemical factory, e.g., Merck-Darmstadt, and, after spectroscopic testing, further investigate them in terms of nuclear physics at Heisenberg's institution. By Joos.~\cite{g46joos}

\subsection{Karl-Heinz H\"ocker}
\subsubsection{Excerpt: ``Dependence on the Energy Output of the Uranium Pile on the Density of the  Uranium and the Density of the Moderator,'' G-41 (April 20, 1940)}\label{tr:g41hocker}

When examining the performance of various uranium machines, it turned out that the greater the density of the substances used, the greater the energy gain from the machine. This general fact is most important when comparing the performance of a uranium machine in which the uranium is in the form of uranium oxide and another in which metallic uranium is used. The density of cast metallic uranium is about 18 [g/cm$^3$] and that of uranium oxide about 3, of which only 2.6 is the uranium in the oxide, such that the densities of metallic uranium and oxide differ by a factor of 7. Naturally, such differences are not to be expected by compressing the scattering substance (coal) or by replacing the water with paraffin, which contains more H atoms per unit volume than water. Nevertheless, this will also be investigated in the second part of the work. Then, the question of whether a denser carbon with some impurities is more favorable than less dense but completely pure carbon will be easily decided.

In the first part, we turn to the question of the additional energy gained by using metallic uranium instead of uranium oxide. This question is of particular interest because the technical production of large quantities of metallic uranium is still difficult at the moment. Since the [neutron] gain from uranium metal is different depending on whether H$_2$O, D$_2$O, or carbon is used as a braking substance, individual tests can estimate whether the procurement of the metal is absolutely necessary, or whether the desired effect can already be achieved with oxide.

Another conclusion is that using metal does not allow a machine to generate energy if it does not already with U$_3$O$_8$. The higher density of the metal therefore only causes the heat energy to be released at a faster rate. For a heating machine, this means that considerably larger amounts of energy can be extracted. 

The calculations are for neutron multiplication in infinite plates. A quantitative extrapolation from infinite to finite plates is not possible. On the other hand, it is self-evident that a finite apparatus will not only be enlarged in thickness. Rather, considerably more layers and much larger surface area will be required for the plates. This means that much larger quantities of braking substance are also required, which in turn creates new difficulties, especially when using D$_2$O. A detailed calculation of a finite apparatus using U$_3$O$_8$ and light water as a braking agent is planned for the near future. There will also be an opportunity to discuss the influence of oxygen in U$_3$O$_8$.~\cite{g41hocker} 

\subsubsection{Excerpt: ``Calculation of Energy Production in the Uranium Pile. Part II. Carbon as a Moderator,'' G-42 (April 20, 1940)}\label{tr:g42hocker}

The result shows that the energy gain [neutron multiplication] from the uranium-carbon machine, with the data assumed here, is an order of magnitude smaller than that from the uranium-D$_2$O machine. It must be noted that the effective cross sections for carbon are quite imprecisely known; in particular, the capture cross section of $\sigma_\text{Cr}=0.3\cdot10^{-26}$ cm$^2$ is only estimated.

It turns out that even for very small deviations from the value of the capture cross section assumed above, the uranium-carbon machine no longer produces any neutron multiplication. [The neutron multiplication] $\bar{\nu}=0$ for $\sigma_\text{Cr}=0.78\cdot10^{-26}$. In the opposite direction, even for $\sigma_\text{Cr}=0$, no neutron multiplication greater than 175 per second can be achieved. In no case is the energy production in the uranium-graphite machine comparable to that in the D$_2$O machine. However, since carbon is a substance that is certainly easier to obtain in large quantities than heavy water, the dimensions of a finite machine are nevertheless of interest.~\cite{g42hocker} 

\subsection{Excerpt: Walther Bothe, ``Diffusion Length of Thermal Neutrons in Carbon,'' G-12 (June 5, 1940)}\label{tr:g12bothe}
[p.1] The advantages that carbon offers as a moderating substance over D$_2$O, assuming that the capture cross section is sufficiently small, were highlighted in an earlier report. The capture cross section, together with the scattering cross section, is incorporated in Heisenberg’s theory of the machine as the ``diffusion length $\ell$.'' Hence, the task arises to determine $\ell$ experimentally. This is best done using spheres whose radius is at least approximately equal to the diffusion length, which in the case of carbon is about 70~cm, assuming a capture cross section of $0.003\cdot10^{-24}$~cm. Such large spheres of homogeneous carbon are not easily manufactured. Therefore, in order to initially ascertain whether carbon can even be considered as a braking substance, an approximate determination of $\ell$ was attempted using a cube of carbon measuring 70~cm along each edge, with the neutron source placed at its center.

[p.3] The inner part of the 70~cm graphite cube was made of neatly machined plates, while the outside consisted of coarsely machined blocks of electrographite. It rested in the middle of the room on a light wooden frame. The graphite was not completely homogeneous in density, with an average density of 1.48 [g/cm$^3$]. It contained 0.5\% impurities which, based on spectral analysis by Professor Sch\"uler, consisted of Ca, Ti, V, and Mg; Ca is the most prevalent; boron could not be detected. A 10~cm cube of paraffin was embedded at the center of the graphite block, which could be fully enclosed on all sides by a 0.4~mm Cd sheet. In the center was the Rn+Be source with an initial activity of 90~mCi.

% [p.6] If the two measuring 

[p.7] However, one must still take into account that the detected impurities in the graphite (Ca, Ti, V, Mg) all have appreciable capture cross sections, which makes $\ell$ appear too small. For pure carbon, therefore, the diffusion length could be substantially greater than 70~cm, and the capture cross section smaller than $0.003\cdot10^{-24}$~cm.

In any case, it would now seem worthwhile to carry out further experiments with carbon. The most sensible course of action appears to be to immediately assemble a machine from Preparation 38 [U$_3$O$_8$] and carbon (cf. the earlier report). For instance, pure carbon could be produced by the method proposed by Joos~\cite{g46joos} and then brought to a density of about 1.5 using the electrographite process [i.e., by graphitizing pyrolytic carbon]. 

Heidelberg, Institute of Physics at the Kaiser Wilhelm Institute for Medical Research, 5 June 1940.~\cite{g12bothe} 

\subsection{Excerpt: Wilhelm Hanle, ``Investigation of Cadmium Content of Carbon,'' G-35 (1940)} \label{tr:g35hanle}

[p.1] After it had been demonstrated (cf. Report H11/50 from March 29, 1940~\cite{g46joos}) that the most difficult impurity in carbon to remove---boron---is present in sugar only in an inconsequential amount, sugar was now examined for its cadmium content. Since the capture cross section of cadmium is $4000\cdot10^{-24}$~cm$^2$ compared to boron’s $500\cdot10^{-24}$~cm$^2$, the purity requirements are even stricter regarding cadmium. If one stipulates that the impurities contribute less than one third to the capture cross section of the coal and takes the carbon cross section to be $0.003\cdot10^{-24}$~cm$^2$ (cf. Heisenberg’s calculation and Bothe’s report), then one must require the cadmium fraction to be less than $2.5\cdot10^{-7}$.

The investigation of cadmium content was conducted on both sugar produced in-house and that obtained from Schering-Kahlbaum. The strong resonance line at 2288 {\AA} was used for cadmium detection... The sugar was powdered and examined spectrally. A copper arc served as the light source. Ordinary copper, however, is unsuitable for sensitive cadmium detection because it itself contains traces of cadmium. Only copper from Heraeus, deoxidized with beryllium, proved to be cadmium-free.

[p.2] The spectra showed weak but clear cadmium lines, as did the spectrum of the pure sugar. Thus, sugar contains traces of cadmium---somewhat less in the sugar produced in-house than in that obtained from Schering-Kahlbaum. By adding $3\cdot10^{-7}$ atomic fractions of cadmium to the coal, the cadmium line at 2288 could still be noticeably intensified. An addition of $1\cdot10^{-7}$ cadmium to the coal produced no or only a slight intensification of the cadmium line. The proportion of cadmium in the sugar was therefore on the order of $10^{-7}$ and certainly less than $3\cdot10^{-7}$. This level of purity just suffices for experiments with slow neutrons. However, the fact that sugar contains cadmium at all likely makes it necessary to test any carbon used for neutron experiments for cadmium beforehand and, when producing larger quantities of carbon for neutron experiments, to continuously monitor the cadmium content.~\cite{g35hanle} 

\subsection{Walther Bothe and Peter Jensen}\label{s:3jensen}
\begin{comment}
    In MCNP, the neutron volume flux through each indicator was directly tallied (F4 tally). B\&J's data are available in the translation, \S\ref{tr:g71bothejensen}. Simulating $10^7$ neutrons per second, all MCNP results had a standard error of about 2\%. 

Following B\&J's notation, we denote $\phi'$ as the flux with the Cd shield and $\phi$ as the flux without. From the MCNP data, the flux ratios are:
\begin{align}
    \frac{\phi_{50}^\text{slow}}{\phi_{6}^\text{slow}} = \frac{\phi_{50}-\phi_{50}'}{\phi_{6}-\phi_{6}'} &= \boxed{1.35 \pm 0.01}, \nonumber\\
    \frac{\phi_{55}^\text{slow}}{\phi_{50}^\text{slow}} = \frac{\phi_{55}-\phi_{55}'}{\phi_{50}-\phi_{50}'} &= \boxed{1.06 \pm 0.01 }, \nonumber
\end{align}
which are similar to B\&J's experimental values of $1.32\pm0.05$ and $1.12\pm0.03$, respectively. B\&J had theoretically predicted the latter ratio to be $1.06$, which is the value obtained from MCNP. 

The diffusion length $L$ the comes out to $35.8\pm0.6 \text{ cm}$. Using the modern scattering mean free path $\lambda_s$ and the graphite atom density $n$ at 1.7 g/cm$^3$, the MCNP--calculated slow capture cross section of the Siemens electrographite is:
\begin{align}
    \boxed{ \overline{\sigma}^\text{slow,MCNP}_a = 7.92\pm0.27 \text{ mb} }, \label{eq:xsBJ}
\end{align}
which is identical to the 7.9 mb value calculated from B\&J's data prior to ashing. 
\end{comment}

\subsubsection{``The Absorption of Thermal Neutrons in Electrographite,'' G-71 (January 21, 1941)}\label{tr:g71bothejensen}
\textbf{1. Goal and Arrangement of the Experiments.} Carbon can be considered as a braking substance for the machine if its absorption cross section $\sigma_a$ for thermal neutrons is sufficiently small. According to Heisenberg’s calculations, with $\sigma_a=3\cdot10^{-27}$ cm$^2$ the machine should just about work. The calculation includes $\sigma_a$ together with the scattering cross section $\sigma_s$ as the ``diffusion length $\ell$.'' Since $\sigma_a\ll\sigma_s$ is certain, the approximation formula can be used for the calculation:
\begin{equation}
    \ell = \sqrt{\frac{1}{3}\lambda_s \lambda_a};\quad \lambda_s = \frac{1}{N\sigma_s};\quad \lambda_a = \frac{1}{N\sigma_a}; \label{eq:g71bothe-1}
\end{equation}
$N=$ number of atoms per cm$^3$. It is therefore feasible to measure the diffusion length directly. 

Earlier measurements (Report ``The Diffusion Length for Thermal Neutrons in Carbon''~\cite{g12bothe}) on rather inhomogeneous graphite of not very high purity ($\sim$1\% ash by weight) had not led to a definitive conclusion. The new measurements were performed with electrographite from Siemens Plania. This is probably the purest carbon that is manufactured in large quantities. We determined the ash content to be 0.09\%; the ash was also specifically examined for neutron absorption (Fig. 3 [not reproduced here]). The material was available in cleanly machined, fairly homogeneous blocks. Another advantage was its relatively high density of 1.7.

This time, the diffusion measurements were carried out using the density measurement method used by Döpel and Heisenberg for D$_2$O. Although this requires more effort in our case, it is probably somewhat more accurate than the measurement method used in the previous study.

From suitably cut graphite blocks, a self-supporting sphere with a diameter of 110 cm was constructed (Fig.~1). Only the lowest layers were slightly supported by a light stand made of aluminum. The sphere rested on a 10 cm high cylinder made of paraffinized wood; a slot was left free in between for the insertion of cadmium. The whole thing was surrounded by a tight-fitting, watertight, open-topped rubber cover and stood on a sturdy iron plate (Fig. 2). A lifting device could be attached to this to lower the sphere into a water container 135 cm in diameter. The thickness of the water was therefore at least 10 cm on all sides; this is the saturation thickness [penetration depth] for thermal neutrons. A Cd layer could be placed between the carbon and the rubber cover. Fig.~3 shows the graphite sphere partially covered with Cd strips, which was held together by Leukoplast [tape].

A vertical square channel was cut out in the graphite sphere for inserting the radiation source and the indicators, which could be refilled with tight-fitting graphite blocks. The neutron source $Q$ was located in the center of the sphere (Fig.~1). Rn-Be with an initial activity of around 70~mCi was used as a source. In order to achieve good consistency, the beryllium powder was pressed in the form of a cruciform and completely melted with glass, except for a filling capillary at the tip of the projectile. (A stronger Ra-Be mixture, which would certainly have allowed for higher measurement accuracy, was unfortunately not yet available to us.) Indicators $I_1$ and $I_2$ were placed 6 and 50~cm, respectively, from the center. For some measurements, a third indicator $I_3$ was placed directly on the side of the graphite. The indicators contain around 24 mg Dy$_2$O$_3$ on a circular area of 2~cm diameter. Their relative sensitivities were specifically determined and found to match the ratio of their dysprosium content. The Dy$_2$O$_3$ was spread on thin aluminum sheets. Control experiments showed that the aluminum’s activity did not cause biases.

\textbf{2. Measurements.} The indicators were left in the graphite sphere overnight and measured the next day for 160 minutes each. For this purpose, they were placed in a well-defined position in front of a thin-walled aluminum counter tube. The sensitivity of the counting tube was constantly monitored with a weak calibration mixture of U+UX. The counters had high resolution (Flammersfeld). The initial activity amounted to a maximum of 200 counts/min with a background of around 8 counts/min. Small corrections were applied for uncounted deflections off the detector ($<2\%$), incomplete activation, any residual activity from a previous irradiation, and slight differences in the indicators’ sensitivities. All measurements in a given series were normalized to the initial source strength. The corrected counts thus obtained represent a measure of the local density of thermal neutrons. Since the scattering path length in carbon is 3 cm, all conditions for a clean density measurement were fulfilled at least for the indicators $I_1$ and $I_2$ (1) [Eq.(\ref{eq:g71bothe-1})]. Indicator $I_3$ served merely to determine whether any boundary effects occurred that might interfere with the measurement when placed too close to the surface of the sphere. 

Let $\rho(r)$ denote the neutron density at a distance r from the center if the carbon sphere is not surrounded by Cd, $\rho'(r)$ the density with Cd; then $\rho-\rho'$ represents the neutron distribution that would be produced by a [thermal neutron] source uniformly covering the surface of the sphere, and this is (Heisenberg):
\begin{equation}
    \rho - \rho' = \frac{\ell}{r}\cdot\sinh\frac{r}{\ell}, \label{eq:g71bothe-2}
\end{equation}
if the [normalization] value at the center is set to one. From this, $\ell$ can be calculated.
\begin{table}[h!] \centering
\caption{Example of a measurement series.}\label{tb:g71bothe-1}
\begin{tabular}{l|rrr}
\toprule
           & \multicolumn{1}{l}{$r_1=6$ cm} & \multicolumn{1}{l}{$r_2=50$ cm} & \multicolumn{1}{l}{$r_3=55$ cm} \\ \midrule
without Cd & $\rho_1=31850$  & $\rho_2=9780$  & $\rho_3=7320$ \\
with Cd    & $\rho'_1=26820$ & $\rho'_2=3112$ & $\rho'_3=\phantom{0}244$ \\ \midrule
difference & 5030            & 6668           & 7076          \\ \bottomrule
\end{tabular}
\end{table}
\begin{equation*}
    \frac{\rho_2 - \rho'_2}{\rho_1 - \rho'_1} = 1.32\pm0.09\qquad \frac{\rho_3 - \rho'_3}{\rho_2 - \rho'_2} = 1.06\pm0.04.
\end{equation*}
Table 1 shows a measurement series. These numbers show the following facts. Without Cd, the neutron density in the graphite drops sharply towards the outside. However, a very large part of this neutron density remains even when the graphite is surrounded by Cd. This part comes from neutrons that are slowed down to thermal energies in the graphite itself; this reflects the considerable braking effect of the graphite. The difference with and without Cd, on the other hand, shows a clear increase [in thermal flux] towards the outside. This difference is due to neutrons that have been slowed down in the surrounding water and diffuse into back into the sphere as thermal neutrons. Here, a noticeable absorption takes occurs. The difficulty of such measurements can be recognized: to obtain a well-measurable amount of absorption, the sphere must have a certain minimum size; yet in that case, the undesired thermal neutrons produced in the graphite itself far exceed those coming from the water. 

Several such measurement series yielded the overall result:
\begin{equation*}
    \frac{\rho_2 - \rho'_2}{\rho_1 - \rho'_1} = 1.36\pm0.05\qquad \frac{\rho_3 - \rho'_3}{\rho_2 - \rho'_2} = 1.12\pm0.03
\end{equation*}
From the first of these two numbers, one can calculate the diffusion length using Eq.(2) [Eq.(\ref{eq:g71bothe-2})]:
\begin{equation*}
    \ell = 36 \pm 2 \text{ cm}.
\end{equation*}
With this value, the theoretical value for the second number is 1.06, in close agreement with the measured number. Thus, if the density distribution near the surface deviates at all from the theoretical expectation, these edge effects are at any rate so weak that the above value of the diffusion length cannot be significantly distorted by them, especially since the outer indicator $I_2$ was almost twice the scattering path length away from the surface.

With the values $\sigma_a=0.003\cdot10^{-24}$; $\sigma_s=4\cdot10^{-24}$, on which Heisenberg's calculation is based, the diffusion length $\ell_0=61$ cm would be expected. The measured diffusion length is so much smaller, and thus the absorption so much stronger, that \textbf{the graphite examined here should hardly be considered as a braking substance for the machine} [emphasis in original].

\textbf{3. Examination of the ash.} The next question is whether the observed additional absorption is attributable to the carbon itself or to some impurity. To investigate this, the ash was examined. A representative sample of 3.6~kg of the graphite was carefully incinerated in a platinum furnace, leaving 3.3~g of ash. This ash was then pressed into a disk 2.4 cm in diameter and 0.6~cm thick. The absorption coefficient of this disk for thermal neutrons was measured. The indicators contained 250 mg Dy$_2$O$_3$ per cm$^2$. The ash disk was placed on the indicator so that about 1/4 of the scattered radiation was measured. The neutron source was Rn+Be in a 10~cm paraffin cube placed 10~cm from the indicators. The measured absorption, corrected for oblique beam path, amounted to 12.5$\pm$1.5\%. From this the scattered absorptions are deducted, which the company's ash analysis estimates at 5.5\%. Thus, about 7\% remains as the true absorption.

We now ask what absorption coefficient of the ash would be expected if all the impurities that reduce the diffusion length from the assumed theoretical value $\ell_0=61$ cm to the measured value $\ell=36$ cm were contained in the ash. From Eq.(1) [Eq.(\ref{eq:g71bothe-1})] one can easily obtain the relationship:
\begin{equation}
    \frac{1}{\ell^2}-\frac{1}{\ell^2_0} = 3N_C\sigma_{sC}\sum N_V \sigma_{aV}, \label{eq:g71bothe-3}
\end{equation}
where the indices $C$ and $V$ refer to the carbon and the impurities, respectively; the contribution of the impurities to the scattering is neglected. With $\sigma_{sC}=4\cdot10^{-24}$ this yields:
\begin{equation*}
    \sum N_V \sigma_{aV}=4.9\cdot10^{-4}\text{ cm}^{-1}.
\end{equation*}
The packing density of the carbon is 1.68, the ash weight $0.092\%$, therefore the absorption cross section sum should be for 1~g of ash:
\begin{equation*}
    \frac{4.9\cdot10^{-4}}{1.68\cdot9.2\cdot10^{-4}} = 0.32\text{ cm}^2/\text{g}
\end{equation*}
Our ash compact had 0.73 g/cm$^2$, so it should have absorbed $21\%$. However, this does not account for the fact that a different mean neutron velocity is effective in the direct absorption measurement than in the diffusion measurement. In the absorption measurement with an ideal $v^{-1}$ indicator, the effective velocity is $0.89\sqrt{2kT/m}$, whereas in the diffusion measurement with a light-nucleus, weakly absorbing medium it is $1.13\sqrt{2kT/m}$, i.e., just twice as large (Report of Bothe and Flammersfeld, ``The Cross Sections of U-238 for Thermal Neutrons from Diffusion Measurements''---this does of course not apply to the highly absorptive atoms that do not obey the $1/v$ law, such as Cd). Due to the $1/v$ law, we should have obtained about $25\%$ absorption instead of the measured $7\%$. \textbf{Thus, only a small part of the shorter diffusion length can be attributed to impurities in the graphite} [emphasis in original].

The measured absorption of $7\%$ can be fully explained by the analytically detected elements. Among these, calcium plays the largest role with a total cross section of $11\cdot10^{-24}$. Strongly absorbing elements such as Cd are apparently not present in disruptive quantities.

\textbf{4. Result.} After the absorption measurement on the ash, we can now correct for the detected impurities using Eq.(3) [Eq.(\ref{eq:g71bothe-3})] and obtain the diffusion length in carbon with the density 1.68:
\begin{equation*}
    \ell_0 = 39\pm2\text{ cm};\quad \ell_0\cdot\text{density}=66\pm3.4\text{ g/cm}^2.
\end{equation*}
With the scattering cross section of $4\cdot10^{-24}$, the absorption cross section of carbon is calculated according to Eq.(1) [Eq.(\ref{eq:g71bothe-1})]:
\begin{equation*}
    \boxed{\sigma_a = (7.5\pm1)\cdot10^{-27}\text{ cm}^2},
\end{equation*}
for a neutron velocity of $1.13\sqrt{2kT/m}$.

\textbf{5. Discussion.} The stated values can hardly be significantly distorted by any undetected impurities in the graphite. During incineration, the temperature did not come close to the very high temperatures to which the graphite is exposed during its production. Therefore, aside from those remaining in the ash, the only other possible impurities could be combustible, e.g., hydrocarbons, and this is very unlikely. In any case, based on the current state of theory, it can be concluded that carbon, even if it is produced with the best-known technical processes and kept completely free of mineral impurities, can hardly be considered as a braking substance for the machine, unless the isotope 235 is enriched.

The stated absorption cross section is not inconsistent with other results. Frisch, Halban, and Koch, based on completely different measurements, give an upper limit of $10^{-26}$ for approximately the same velocity. The CO$_2$ measurements by Harteck et al. can also be reconciled with our result if oxygen absorbs far more weakly than carbon, which is supported by the D$_2$O measurements of D\"opel and Heisenberg.

Institute for Physics at the Kaiser Wilhelm Institute for Medical Research, Heidelberg, 20 January 1941~\cite{g71bothejensen} 

\subsubsection{Excerpt: ``The Absorption of Thermal Neutrons in Carbon,'' \textit{Zeitschrift f\"ur Physik} \textbf{122}(12) 749--755 (September 1944)}
% \begin{center}
% With 2 figures. (Received January 21, 1944.)
% \end{center}
\textit{Abstract.} The diffusion length of thermal neutrons in electrographite was measured and the absorption cross section calculated. The results were corrected for pure carbon by determining the absorption of the ash separately. (This investigation was completed in January 1941. Publication was delayed for external reasons.)

\textit{6. Evaluation and Result.} For the scattering cross section of the C nucleus for thermal neutrons, values between $4.1$ and $4.8\cdot10^{-24}$ cm$^2$ have been measured (J. R. Dunning et al., Phys. Rev. \textbf{48}, 265, 1935 \cite{dunning35}; M. Goldhaber and G. H. Briggs, Proc. Roy. Soc. London (A) \textbf{162}, 127, 1937; M. D. Whitaker and W. C. Bright, Phys. Rev. \textbf{60}, 155, 1941.). We will use $4.5\cdot10^{-24}$ cm$^2$. At a packing density of 1.68, this yields $\lambda_s=2.6$ cm. With the diffusion length 36 cm found here, according to equation (2) [Eq.(\ref{eq:bothe44-2})],
\begin{equation}
    \ell = \sqrt{\frac{1}{3}\lambda_s \lambda_a};\quad \lambda_s = \frac{1}{N\sigma_s};\quad \lambda_a = \frac{1}{N\sigma_a} \label{eq:bothe44-2}
\end{equation}
the absorption path length $\lambda_a = 1480$ cm and the absorption cross section $\sigma_a = 8\cdot10^{-27}$ cm$^2$ would be calculated. However, part of this is due to impurities. The investigated ash disk had an area of 4.15 cm$^2$ and absorbed 7\%, so the net absorbing cross section of the ash  was $4.15\cdot0.07 = 0.29$ cm$^2$. This value is per 3.6 kg graphite that was ashed. Per carbon atom, therefore, is $1.6\cdot10^{-27}$ cm$^2$. Subtracting this from the above value $\sigma_a$ yields the average absorption cross section of the pure C atom:
\begin{align*}
    \sigma_a = (6.4\pm1)\cdot10^{-27}\text{ cm}^2.
\end{align*}
(As K. Diebner informs us, the incorrect transcription of this result in \textit{Physikalische Zeitschrift} \textbf{43}, 440, 1942 \cite{diebner42} was probably due to a typographic error.\footnote{In his 1942 compendium of known neutron data (a successor to Weizs\"acker's 1939 tables~\cite{weizsacker}), Diebner mistakenly wrote carbon's thermal absorption was 3~mb ``based on a private communication from W. Bothe,'' when that was simply the theoretical value Heisenberg computed~\cite{diebner42}.}) Whether this absorption belongs to the $^{12}$C or $^{13}$C nucleus cannot be decided for the time being. In the second case, the absorption cross section of the $^{13}$C nucleus would be $0.6\cdot10^{-24}$ cm$^2$.

The diffusion length for pure carbon is calculated to be $\ell = 40$~cm according to (2) [Eq.(\ref{eq:bothe44-2})] with the corrected $\sigma_a$, or:
\begin{align*}
    \ell \cdot \text{density} = (67\pm4)\text{ g/cm}^2.
\end{align*}
There is still a small possibility that not all impurities, such as tiny amounts of boron, were captured with the ash, although incineration did not nearly reach the high temperature to which the material was already exposed during the manufacturing process. The specified absorption cross section could therefore be rather too high and thus the diffusion length rather too small.

Frisch, Halban, and Koch (\textit{Royal Danish Academy of Sciences} XV, No. 10, 1937) have found, according to a fundamentally different method, an upper limit of $10^{-26}$ cm$^2$ for the absorption cross section of the C nucleus. This is in agreement with our result.

\textit{Summary.} For the absorption cross section $\sigma_a$ and the diffusion length $\ell$ of thermal neutrons in pure carbon, the following values were measured:
\begin{align*}
    \sigma_a=(6.4\pm1)\cdot10^{-27}\text{ cm}^2;\qquad\ell\cdot\text{density}=(67\pm4)\text{ g/cm}^2.
\end{align*}
\textit{Heidelberg}, Institute for Physics at the Kaiser Wilhelm Institute for Medical Research, 14 January 1944.~\cite{bothe44}

\subsection{Wilhelm Hanle}
\subsubsection{``On the Detection of Boron and Cadmium in Carbon,'' G-85 (April 18, 1941)}\label{tr:g85hanle}

If carbon is used as a braking substance in the uranium machine, care must be taken to ensure that the boron and cadmium content is sufficiently small because of their large capture cross section for slow neutrons. The majority of carbons contain a considerable amount of boron, as shown in the table [Table \ref{tb:hanle}]. The values given were determined spectroscopically. Only the carbons produced from sugar are practically boron-free. The content of the other carbons was determined by comparison with sugar to which a known amount of boric acid had been added. With this method, however, it is possible that only a lower limit is obtained. In graphite, for example, the boron is present as boron carbide and is difficult to evaporate, while the boron from sugar soaked in boric acid evaporates more easily. 
The experiments were carried out by filling finely powdered carbon into the bore of a copper electrode. In the electric arc, some of the carbon sprayed out, and it is possible that the boron from the boric acid completely evaporated while boron from boron carbide only partially vaporized. Indeed, Goldschmidt previously found slightly larger amounts of boron in coal ash. Our method has the advantage of not requiring the coal to be ashed. It provided initial proof that ordinary coal is certainly unusable due to its large boron content, but that sugars are largely boron-free. More recently, the Siemens electrographite used by Bothe in his determination of neutron absorption in carbon was also examined. For the time being, the boron content could only be constrained within two limits: it is less than $10^{-5}$ and certainly greater than $2\cdot10^{-6}$. 
The question now is, can part of the neutron absorption measured by Bothe be caused by the boron content? Bothe measured a neutron absorption cross section of $0.0075\cdot10^{-24}$~cm$^2$, while Heisenberg calculated $0.003\cdot10^{-24}$~cm$^2$. Let us assume that the entire difference of $0.004\cdot10^{24}$~cm$^2$ [\textit{sic}] is due to absorption in boron. Then the boron fraction would have had to be $9\cdot10^{-6}$. The preliminary tests indicate less than $10\cdot10^{-6}$ and greater than $2\cdot10^{-6}$. So it is at least possible that the large neutron absorption in Siemens electrographite comes from boron contamination. Of course, now that the exact boron content matters, the method will be improved by adding boron nitride, which is more difficult to evaporate, to the graphite instead of boron oxide.

Bothe found that the ash of electrographite absorbed about three times less than what is calculated from the absorption before ashing. Thus, if the neutron absorption had been caused by boron, then at least some would have been partially lost during incineration. This is possible because there is an easily vaporizable boron monoxide.

Cadmium has a 10 times larger capture cross section than boron. So, the cadmium measurement was worked out very carefully. It was found that the cadmium content in sugar is on the order of $10^{-7}$ and certainly smaller than $3\cdot10^{-7}$. $10^{-7}$ parts of cadmium would contribute $0.0004\cdot10^{-24}$ to the absorption cross section of graphite, i.e., only about one-tenth of the cross section calculated by Heisenberg for carbon. In the electrographite used by Bothe, the cadmium content is also less than $3\cdot10^{-7}$. Hanle.~\cite{g85hanle} 

\begin{table}[h!] \centering
\caption{Boron content in atomic parts.}\label{tb:hanle}%
\begin{tabular}{@{}lc@{}}
\toprule
Bituminous coal & $10^{-5}$--$10^{-6}$  \\
`` '' ash per Goldschmidt & $3\cdot10^{-4}$  \\
Beechwood charcoal & $10^{-5}$  \\
`` '' ash per Goldschmidt & $10^{-4}$  \\
Graphite from Acheson Co. & $10^{-4}$--$10^{-5}$  \\
Arc lamp carbon, heat-treated & $10^{-4}$  \\
Activated charcoal, Merck & $10^{-5}$  \\
Blood charcoal, powdered & $10^{-5}$  \\
\midrule
Carbon from rock candy    & $<10^{-6}$  \\
`` '' table sugar    &  $<10^{-6}$ \\
`` '' dextrose    & $<10^{-6}$ \\
`` '' potato starch & $<10^{-6}$ \\
Schering sugar & $<10^{-6}$ \\
\midrule
Siemens electrographite & $<10^{-5} >2\cdot10^{-6}$ \\
\bottomrule
\end{tabular}
% \footnotetext{Source: This is an example of table footnote. This is an example of table footnote.}
\end{table}

\subsubsection{Excerpt: ``Spectroscopic Analysis of Carbon, Aluminum, and Beryllium,'' G-153 (March 17, 1942)} \label{tr:g153hanle}

[p.162] The graphite used by Bothe was examined for boron content, as well as the aluminum used by D\"opel and Heisenberg [G-75 or G-136] and the beryllium oxide used by Haxel and Volz [G-91] for boron and cadmium content... As explained in an earlier report~\cite{g85hanle} and at the first conference on P38 [U$_3$O$_8$], particular attention must be paid to traces of boron and cadmium when checking the extent to which neutron experiments can be influenced by impurities.

\underline{Carbon.} The experiments were carried out as follows: The substance was pulverized and placed in a cavity of about 2~mm diameter and 2~mm depth in a copper rod of 5~mm diameter and vaporized in an electric arc. The copper itself contained traces of boron. This and the thick base of the arc affected the accuracy of the measurement. Boron-free copper from Heraeus was used in the earlier investigations. Unfortunately, this copper was used up and it was impossible to obtain new, highly purified copper from Heraeus, so copper from an ordinary workshop had to be used.

When graphite powder was introduced into the cavity of the copper electrode, the boron lines were significantly strengthened. In order to be able to quantify the boron content, measurements were made with reference standards. These consisted of carbon that was as boron-free as possible and to which boron was added in known ratios... The images were captured using a small Zeiss photometer, namely: 1. the carbon line 2479 [{\AA}]; 2. the two boron lines 2497 and 2498; 3. the background to the left and right and in the middle between the boron lines.

[p.163] % A Hansen's stepped aperture [Hansen'sche Stufenblenden-Aufnahme] was used to photograph and photometrize each image, with the help of which the blackening lines of 1, 2, and 3 were drawn and converted into intensities. The background at the location of the boron lines was then approximately determined by forming an average value from the background on each side and in the middle between the boron lines. This value was subtracted from the measured boron line intensity. The intensity of the boron lines was always very weak compared to that of the carbon line, as very small boron concentrations were used. So, either the boron lines were underexposed or the carbon lines were overexposed. They could therefore never be in the favorable linear blackening range at the same time. Thus, the boron lines were exposed normally and then the too strong carbon intensity was weakened by an attenuator such that it also came into the normal blackening range. For this purpose, half of the slit was covered by a thin cellophane skin, which attenuated the carbon line to just the right intensity. Two spectra were then obtained, one below the other, which were exposed at the same time. On one spectrum the boron lines were in the favorable blackening range and on the other spectrum the carbon line. First, it was qualitatively determined that the boron content of the graphite sent to me by Bothe was certainly greater than $10^{-6}$ and less than $10^{-5}$ (atomic parts boron to atomic parts carbon). Then mixtures of boron-free carbon with boron were prepared in the ratio of $1\cdot10^{-6}$, $3\cdot10^{-6}$, $4\cdot10^{-6}$, $5\cdot10^{-6}$, $6\cdot10^{-6}$, $7\cdot10^{-6}$, $8\cdot10^{-6}$ and $9\cdot10^{-6}$ atomic parts carbon to atomic parts boron.
Sugar was used as the carbon reference. Earlier (see the earlier report and the first conference on P38) it had been stated that carbohydrates was practically boron-free. At that time, the experiments in G\"ottingen were carried out both with carbon produced from rock candy available there and with activated charcoal purchased from Merck. The samples used at that time had been used up in the meantime. Unfortunately, when the investigations were continued in Giessen, it turned out that the carbohydrates apparently did contain boron in small quantities. This was the case with both the charcoal recently purchased from Merck and with the rock candy bought in Giessen. Initially, the new finding was explained by the fact that the G\"ottingen tests were too qualitative and therefore the low boron content of the sugar had escaped detection at the time. In the case that boron content is not negligible in carbohydrates, an attempt was first made to determine the boron content by adding proportional quantities of boron until a doubling of the boron intensity occurred. This boron content in the sugar was then taken into account in all measurements. This method was not only quite laborious and time-consuming, but also quite inaccurate, as the boron content in sugar obviously fluctuated greatly, resulting in large uncertainties in the boron content of Bothe's graphite determined in this way. Only after many investigations was it possible to find a carbon which, like the sugar previously purchased in G\"ottingen and the activated charcoal previously produced by Merck, contained practically no boron. This sugar was supplied by the company Pfeifer \& Langen, at Elsdorf in the Rhineland, and was described by this company as particularly pure. The carbon produced from this sugar could now serve as a reference substance.

[p.166] The boron content lies between $0.5$ and $1\cdot10^{-5}$ and quite certainly between $0.6$ and $0.8\cdot10^{-5}$ atomic parts of boron per atomic part carbon. Determining such small boron concentrations is probably at the limit of what is measurable. According to the chemists, a chemical determination of such tiny amounts of boron is impossible. However, there is a way forward: one can extract the boron from the graphite using a method developed for soil analysis, producing boron-free graphite. Then, by adding specific amounts of boron and using the same baseline for both sample and reference, it might be possible to eliminate at least this source of error. However, this would only be worthwhile if carbon is really still considered for the [moderation] problem in the current state of nuclear physics investigations. In any case, part of the neutron absorption in the graphite used by Bothe can be explained by boron contamination. Giessen, on 17.III.42. Hanle~\cite{g153hanle} 

\begin{comment}
\subsection{Excerpt: Josef Kremer, Doctoral Dissertation, G-104 (December 1941)}\label{tr:g104kremer}

Some substances are characterized by particularly high cross sections for the absorption of slow neutrons. These elements are: Eu, Sm, Dy, Gd, B, and Cd. Their capture cross sections for neutron absorption are on the order of magnitude of the geometrical cross sections defined in the usual way for fast neutrons, whose energy ranges from about 0.1 to several MeV, though they can assume capture cross sections up to ten thousand times larger for thermal neutrons.

For the capture process, Bethe has given a formula that expresses the functional relationship between the absorption cross section and the neutron energy. It is valid in the range of slow neutrons (energies up to 100 eV)... The two factors of the Bethe relation generally define two main capture regions: one lies at low neutron energies in the thermal region, and the other lies at an energy characteristic of the irradiated element, the resonance region. For an element, if the maximum resonance lies very close to the thermal region, and the two capture regions overlap, then a particularly large capture cross section for thermal energies is observed. This case applies to the substances to be investigated here. The only exception is boron [among the elements studied].~\cite{g104kremer}    
\end{comment}

\subsection{Excerpt: Army Ordnance Office, ``Energy Production from Uranium,'' pp.14, 21, 87--89 (1942)}\label{tr:energie}
[p.14] \textbf{b. Braking Substances.} As long as isotope separation has not been achieved, heavy water (D$_2$O) is considered as the primary braking substance for an energy-producing uranium machine. It is contained in ordinary water in a highly diluted state (1/6000) and must be extracted from that. At the beginning of the war, D$_2$O was produced in only one place in the world: at Norsk Hydro in Rjukan, Norway, in small quantities (20 liters per year).

Extensive experimental work was conducted on producing D$_2$O in Germany. Initially, efforts focused on transferring the Norwegian electrolytic process to Germany. However, this was not feasible on a large scale due to the lack of cheap electricity. Later, the Norwegian facility was expanded to produce one ton per year. Further expansion to about 4--5 tons per year using newly developed leaching processes is in preparation.

Completely new processes have been developed and tested for the production of large quantities of heavy water in Germany, which can be connected to hydrogenation plants, power stations, etc. without affecting their production. The aim of these processes is either to filter out heavy water from ordinary water catalytically or to fractionate it using waste energy [to about 1--10\% heavy water]. Additional processes are then used to achieve higher concentrations. This is a technically and economically feasible way of producing large amounts of heavy water in Germany, many times the Norwegian production.

If large-scale uranium isotope separation succeeds, the problem of braking substances can change entirely. Then, for example, ordinary water and carbon could also be suitable.

[p.21] The braking effect of substances decreases with increasing atomic weight. All light substances up to C were tested for their suitability. Li, Be (but see Appendix II), and B are hardly suitable as they undergo nuclear transformations even with relatively slow neutrons. Carbon has also proved to be unsuitable according to recent measurements. 

[p.87] \textbf{g. The Examination of Braking Substances.} The measurements conducted on this topic focus on diffusion lengths. Since scattering cross sections can mostly be obtained from the literature, the absorption cross section can be calculated from them. We always include [the scattering cross section] below because the test materials sometimes exhibited significant variations in density.

\underline{1) Hydrogen.} Currently, the best values for the cross sections are [in barns, b]:
\begin{equation*}
    \sigma_a = 0.25 \qquad \sigma_s = 48.
\end{equation*}
Hydrogen is characterized by an extraordinarily large scattering cross section, which drops sharply at higher neutron energies. In the form of water, it would be the ideal braking substance. Unfortunately, due to its relatively high capture cross section, it is not suitable for our purpose.

\underline{2) Deuterium.} Measurements by D\"opel and Heisenberg (D1) yielded $\ell = 90$ cm. Assuming a scattering mean free path of 1.7 cm, this results in:
\begin{equation*}
    \sigma_a = (1.1 \pm 2.3)\cdot10^{-27}\text{ cm}^2
\end{equation*}
The construction of a machine using deuterium as a braking substance is currently in preparation and appears promising. [...]

\underline{4) Carbon.} Measurements by Bothe and Jensen (B9) on electrographite with density of 1.7 yielded $\ell = 36 \pm 2$, with $\sigma_s = 4$:
\begin{equation*}
    \sigma_a = (7.5 \pm 1)\cdot 10^{-27}\text{ cm}^2
\end{equation*}
With this capture cross section, a machine with carbon is not possible. Since a detailed examination of the electrographite used later revealed a small boron content, the true capture cross section is probably smaller. However, since it is practically impossible to produce carbon with a higher degree of purity than what was used, it would hardly be suitable as a braking substance.

[p.89] \textbf{h. Summary.} Studies have concluded that heavy water and beryllium could be considered as braking substances. It will be a question of how the machine is used as to which of the two substances is ultimately chosen.~\cite{hwa42}

\subsection{Excerpt: Wolfgang M\"uller, \textit{History of Nuclear Energy in the Federal Republic of Germany}, p.30 (1990)} \label{tr:mueller}

%\textbf{The Moderator}
% In the search for a suitable substance as a moderator with a low atomic weight and low neutron capture, theoretical considerations led to three substances that seemed particularly suitable: water, carbon in the form of graphite, and beryllium, which was soon eliminated from the choices (also in the USA) due to its metallurgical problems. In the water, the oxygen atom (O) fulfilled expectations. However, the two hydrogen atoms (H) in the water molecule (H$_2$O) capture too many neutrons. The heavy isotope of hydrogen, deuterium (D), which occurs in normal water in a ratio of 1:6,500, behaves much more favorably. Because the mass of D is twice that of H, in this case the two isotopes have different properties (which is why the heavy isotope is given its own name, deuterium). This meant that they could be separated using simpler methods than the uranium isotopes, even though the methods in question were still very complicated and energy--intensive. 

Graphite was the first choice as a moderator, although a reactor with a heavy water moderator required significantly lower quantities of uranium and moderator substance, as D\"opel and Heisenberg were able to show. Before the war, heavy water was only needed in very small quantities for laboratory requirements and was essentially only produced by the Norsk Hydro company in Norway, of which the French had a stake, and the Lonza company in Switzerland. A gram cost half a dollar.

However, measurements carried out at Bothe's institution in Heidelberg revealed a much stronger neutron capture [in graphite] than had been expected. The German physicists concluded from this that graphite was not suitable as a moderator and decided to use heavy water. However, this was a mistake. In the USA, Fermi and Szilard had established in 1940 that graphite was indeed suitable.

Although the production of large quantities of ``nuclear-pure'' graphite also initially caused difficulties in the USA, Fermi and his team were able to build their reactor in the hall under the stands of a sports stadium near Chicago, in which the first nuclear fission chain reaction in the world was set in motion on December 2, 1942, using 350 tons of graphite. Obviously, the excessively high German values had been caused by impurities in the graphite. This was probably also suspected by Bothe, but the graphite was not analyzed in detail anywhere. Because the difficulties in procuring heavy water were largely to blame for the delays in German reactor development during the war, Bothe's measurements were later heavily criticized in some works. However, the dispute seemed to be more academic because the electrographite produced by Siemens Plania Works AG for Carbon Products in Berlin-Lichtenberg, one of the world's leading graphite manufacturers, was the purest that could be obtained in Germany, and it was obviously not suitable as a moderator. However, the former factory director of Siemens Plania wrote to Heisenberg after the war that he was surprised to read the graphite problems from one of Heisenberg's essays, because this question had surprisingly not been addressed to him or his employees: ``There is no doubt at all that it would have been possible for us at any time, for scientific purposes, to produce and deliver electrographite a greater degree of purity than otherwise required for technical purposes.'' In his reply, Heisenberg blamed the bureaucracy for this lack of information: ``You are probably quite right in saying that it would probably have been possible to produce graphite of the necessary purity in Germany, too, if the authorities had been sufficiently interested to take a closer look at this problem. But this `if' was decisive for the entire progress of the work'' (Quote from Dr. Erich H\"ohne to Heisenberg dated September 9, 1947; Heisenberg to H\"ohne dated September 15, 1947. Heisenberg Files, Correspondence 1946/1947).~\cite{Mueller}

\end{appendices}

\bibliographystyle{unsrtnat}
\bibliography{main}  %%% Uncomment this line and comment out the ``thebibliography'' section below to use the external .bib file (using bibtex) .

%%% Uncomment this section and comment out the \bibliography{references} line above to use inline references.
% \begin{thebibliography}{1}

% 	\bibitem{kour2014real}
% 	George Kour and Raid Saabne.
% 	\newblock Real-time segmentation of on-line handwritten arabic script.
% 	\newblock In {\em Frontiers in Handwriting Recognition (ICFHR), 2014 14th
% 			International Conference on}, pages 417--422. IEEE, 2014.

% 	\bibitem{kour2014fast}
% 	George Kour and Raid Saabne.
% 	\newblock Fast classification of handwritten on-line arabic characters.
% 	\newblock In {\em Soft Computing and Pattern Recognition (SoCPaR), 2014 6th
% 			International Conference of}, pages 312--318. IEEE, 2014.

% 	\bibitem{keshet2016prediction}
% 	Keshet, Renato, Alina Maor, and George Kour.
% 	\newblock Prediction-Based, Prioritized Market-Share Insight Extraction.
% 	\newblock In {\em Advanced Data Mining and Applications (ADMA), 2016 12th International 
%                       Conference of}, pages 81--94,2016.

% \end{thebibliography}

\end{document}